 \documentclass[11pt]{article}

 \pdfoutput=1   

\usepackage{graphicx}
\usepackage{amsmath}

\usepackage{enumitem}

\usepackage[francais,english]{babel}

\usepackage{float}
\usepackage{ifthen}
\usepackage{color}

\usepackage{hyperref}

\begin{document}





\title{{\bf The available-enthalpy (flow-exergy) cycle. \\
   Part-I: introduction and basic equations.}}

\author{by Pascal Marquet. {\it CNRM/GMAP. M\'et\'eo-France, Toulouse, France.} \\
 Email: pascal.marquet@meteo.fr}

\date{\today}

\maketitle


\vspace*{-10mm}

\begin{center}
{\em Copy of a CNRM-Note submitted in two parts in April 2001 to the
 \underline{Quarterly Journal of the Royal Meteorological Society}.} \\
{\em Published in Vol.129, Issue 593, Part-I (2445--2466) Part-II (2467--2494), July 2003, Part B.} \\
  Part-I: \url{http://onlinelibrary.wiley.com/doi/10.1256/qj.01.62/abstract} \\
  Part-II: \url{http://onlinelibrary.wiley.com/doi/10.1256/qj.01.63/abstract} \\
Comments and corrections are added in footnotes. 
\end{center}
\vspace{1mm}



\vspace*{-2mm}

\begin{abstract}
A diagnostic package is derived from the concept of
specific available enthalpy, leading to the definition 
of a local and complete energy cycle.
It is useful to understand the transformations of energy
occurring at any particular pressure level or pressure layer of a 
limited area domain.
The global version of this diagnostic tool
is very similar to the cycle of Lorenz, but the local counterpart
contains several additional terms, with zonal, eddy and static-stability 
components close to definitions already given by 
Pearce. 
The new cycle takes into account the flow of energy components
across the vertical and horizontal boundaries, with additional
conversion terms involving potential energy,
leading to accurate computations of dissipation and 
generation terms obtained as residuals. A 
new accurate temporal scheme is proposed in order
to allow use of a large time interval in future
numerical applications.
Finally, comments are made on the arbitrary choice for
two constant reference values for pressure and 
temperature.
\end{abstract}

 \section{\Large \underline{Introduction}.} 
 \label{section_1}

Understanding of energy transformations
occurring in the atmosphere is still a
subject of research, and several methods exist
to investigate observed atmospheric
energetics. Margules (1905) performed
an application to a single column of fluid, and 
its generalization to the general circulation has been
realized by Lorenz (1955, hereafter L55). More recently,
different local versions of the Lorenz cycle have
been published when authors are concerned 
with small-scale phenomena like  tropical or mid-latitude
cyclogenesis, baroclinic-wave development or 
frontal cyclogenesis, all associated with limited-area 
domains (e.g. Muench 1965; Brennan and Vincent 1980;
Michaelides 1987).

Previous local studies based on the
Lorenz method have been able to catch 
the main features of local energy transformations,
including usual baroclinic or barotropic conversions 
and involving classical differential heating terms, with
boundary terms different from zero only in the case 
of limited-area domains.

However, these local versions  lead to certain
inconsistencies. There are two main
problems resulting from global-scale assumptions
that do not hold for limited-area domains.
Firstly, all terms in the energy cycle 
are integrated part by part through the whole
atmosphere and, as
a consequence, finite vertical-extent
domains or an isolated level cannot be considered. 
Secondly, the mean vertical velocity 
$\overline{\omega} \: = \: \overline{d p / d t}\:$,
where $p$ is pressure, 
is supposed to cancel out when it is
averaged over any horizontal layer. However this is
only true for a surface surrounding the whole 
earth, which excludes the use of Lorenz's cycle for
local studies owing to large impacts caused by
these approximations (Saltzman and Fleisher, 1960).
For example, the mean conversion term
$- \: R \: \overline{\omega} \: \overline{T} \: / {p}$
and eddy component 
$- \: R \: \overline{ {\omega}^{\prime} 
\: {T}^{\prime} } \: / {p}$
can be of the same order of magnitude because, even if
$|\overline{\omega}|$ is a tenth of
$|{\omega}^{\prime}|$, $|\overline{T}|$ is
commonly $10$ times greater than $|{T}^{\prime}|$
(Symbols are defined in appendix~A).
As a result, the mean conversion term cannot be neglected 
in limited-area energetics.

These local studies present other unrealistic 
features--for instance when dissipation
and generation terms are the only unknown quantities 
and are computed as residuals of the cycle.
It is often mentioned that these residuals 
are too large and that they lead to unbalanced terms,
like the conversion terms with potential energy.

In other words, there is a need for a new kind of 
local energy cycle without any
missing terms and where approximations would 
be overcome. The method adopted to revisit the approach 
of local energetics in meteorology
is to start with a set of local and
exact equations for temperature and wind, to
define appropriate availability functions,
to specify a reference state and finally
to compute the average values over a given
pressure level for a limited-area domain.
This methodology ensures that
all the terms will be present in the local
version, even if some of these terms cease to
exist in the globally averaged version.

A new local and exact available-enthalpy
 cycle is proposed in this paper.
It will clear up the difficulties 
encountered with previous limited-area
applications and, on the global stage,
will lead to results more usually expected, 
including 
baroclinic and barotropic instabilities.
This new cycle is based on the concept of
available enthalpy described in Marquet (1991,
hereafter M91), following the
proposition of Sir Charles Normand (Normand 1946) when
he chose a direct approach in terms of 
enthalpy (total heat) in place of the total 
potential energy used by 
Margules and Lorenz.

Part~I of this paper was taken from a 
thesis (Marquet, 1994). The concept of 
available enthalpy has not been widely applied 
in meteorology and a short review of its 
development, both in meteorology and in general
physics, will be discussed in the section~2.
Kinetic-energy and available-enthalpy components 
are presented in section~3 and the fundamental
energy equations are used in section~4 to 
define the limited-area available enthalpy cycle.
Associated with this, a new accurate temporal scheme 
is proposed in section 5 to allow future 
use of large time intervals in Part~II,
where applications to idealized 
baroclinic waves will be presented. A discussion 
on the prescribed `reference' pressure and temperature 
is presented in section~6. The final conclusion 
appears in section~7.
Symbols and notations for Parts~I and II are explained in Appendix~A.

 \section{\Large \underline{The energy availability concepts in meteorology and} \\
                 \underline{thermodynamics}.} 
 \label{section_2}

      \subsection{In thermodynamics.} 
      \label{subsection_2.1}

Problems of defining energy availability have been tackled in many ways in physics, and different available-energy and available-enthalpy concepts was developed during early developments in thermodynamics.
The aim was to compute part of the total energy contained in a closed or open system that can be available for useful technical work.\footnote{\color{blue} 
A review in available in Marquet (1991) \url{http://arxiv.org/abs/1402.4610} {\tt arXiv:1402.4610 [ao-ph]}.}
All availability functions introduced by 
Lord Kelvin\footnote{\color{blue} 
The concept of ``Motivity'' has been introduced by W. Thomson when he explored the application of the concept of ``Motive Power of Heat'' defined by Sadi Carnot (1824).
Thomson published the explicit formulae $W = \int_{T_0}^T \: c_p \: (1-T_0/T') \: dT' = c_p \: [ \: (T-T_0) - T_0 \: \ln(T/T_0) \: ] $ in 1853 for defining the maximum work that can be obtained by bringing the uneven temperature $T(x,y,z)$ of all the matter to the constant equilibrium one $T_0$.
This corresponds to what is called ``flowing exergy'' nowadays, namely to: $(H-H_0) \: - \: T_0 \; (S-S_0)$.
}
(Thomson, 1849, 1853, 1879),
Maxwell\footnote{\color{blue} 
The available energy was erroneously called ``entropy'' by Maxwell in first editions of the book (still in the third one in 1872), being influenced by the Scottish mathematical physicist P. G. Tait and  differently from the way Rudolf Clausius (1865) has defined the modern version of this concept.
The formula $(U-U_0) \: - \: T_0 \; (S-S_0)$ was was written explicitly in the next editions of ``Theory of heat'', following the influence of Gibbs (1879).
}
(1871) 
or Gibbs\footnote{\color{blue} 
Gibbs called the quantity $W_{max} = (U-U_0) \: - \: T_0 \; (S-S_0) \: + \: p_0 \; (V-V_0)$ the ``available energy'' of a body.
This is called ``non-flow exergy'' nowadays.
Gibbs also called ``capacity for entropy'' the maximum available work $W_{max} = T_0 \: \Delta S_{tot}$ expressed in  terms of the temperature of the surrounding thermostat at $T_0$ and the change in total entropy of the system $S_{tot}$, where ``total'' means the sum of the change for the body and for the surrounding thermostat at $T_0$ and $p_0$. It is likely that this definition in terms of change in total entropy is the more general one.
}
(1873) 
depend on the local
internal energy ($e_i$), enthalpy ($h$) and 
entropy ($s$) of the fluid. 
A reference state must be specified, generally given
by a constant `reference' temperature ($T_r$) with
associated pressure ($p_r$) and specific volume
(${\alpha}_r$). There is only one available-enthalpy 
function:
\begin{equation}
a_h \; = \; ( h - h_r ) \: - \:  T_r \: (s - s_r) \: ,
\label{equa_ah}
\end{equation}
but two available energies have been defined.
Two functions, one simple and the other one more 
complex, are given by
\begin{equation}
{a}_{e1} \; = \; [ \: e_i - {(e_i)}_r \: ] \: - \: T_r \:  (s - s_r) \: ,
\nonumber 
\end{equation} 
and
\begin{equation}
{a}_{e2} \; = \; [ \: e_i - {(e_i)}_r \: ] 
   \: + \: p_r \: ({\alpha} - {\alpha}_r)
   \: - \: T_r \: (s - s_r) \: .
\nonumber 
\end{equation}

      \subsection{In meteorology.} 
      \label{subsection_2.2}

\begin{figure}[t]
\centering
\includegraphics[width=0.95\linewidth,angle=0,clip=true]{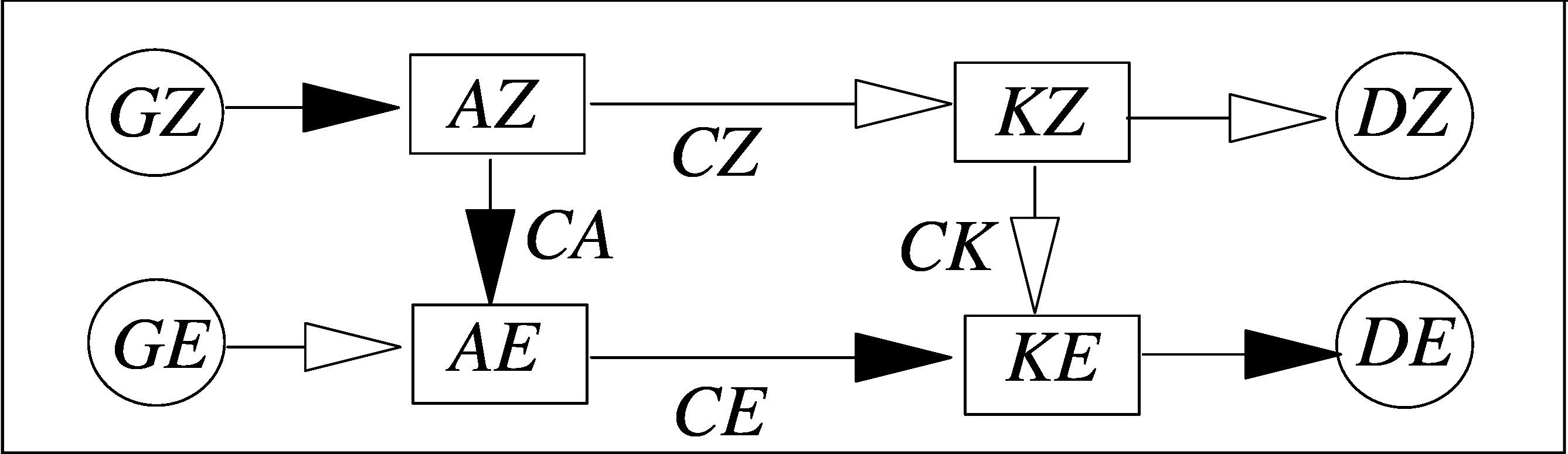}
\caption{
{\it \small 
A diagram for the Lorenz cycle. The baroclinic
         instability is depicted by the dark arrows: 
         $GZ \rightarrow \;$$CA \rightarrow \;$$CE \rightarrow \;$$DE$.}
\label{Fig_1}
\label{FigCLorenz}}
\end{figure}

Without referring to thermodynamic theories, and
following the ideas of Margules, Lorenz defined
available potential energy ($APE$) as the maximum part 
of the sum of internal and potential energies -- also
called total potential energy ($TPE$) -- which can be
transformed under adiabatic motion into kinetic 
energy ($K$).
Margules and Lorenz treated kinetic energy as the
useful energy in the atmosphere because of its
easily visible effects.
Lorenz's approach can be summarized by a
cycle (see Fig.~\ref{FigCLorenz}), where each $APE$ 
and $K$ reservoirs are separated into two parts
($APE=AZ+AE$ and $K=KZ+KE$). 
The four components $AE$, $AZ$, 
$KE$ and $KZ$ correspond to a separation of the
general circulation into zonally symmetric and eddy 
parts (denoted by the suffixes $Z$ and $E$ respectively).
Energy generation terms $GE$ and $GZ$ provide energy 
to $AE$ and $AZ$, with dissipation terms $DE$ and $DZ$ 
acting on $KE$ and $KZ$. 
The terms $CA$, $CE$, $CK$ and $CZ$ are
conversion terms as shown in Figure~{\ref{FigCLorenz}}.

The result obtained by Lorenz demonstrates
the maintenance of the general
circulation ($KE$) by baroclinic instabilities ($CA$ and $CE$).
The horizontal differential heating term $GZ$
supplies energy to the internal reservoir $AZ$ which
is transformed into kinetic energy by 
baroclinic conversions $CA$ and $CE$,
so that the $KE$ component is maintained despite 
the continuous dissipation $DE$.

Apart from the global Lorenz cycle and associated local
versions, other availability functions have been defined in
meteorology. It has been demonstrated in 
Marquet (1995) that most of them are associated with
special thermodynamic availability functions.
For instance, the available 
energy ${a}_{e1}$ corresponds to 
a quantity $T_0 \Sigma$ 
called `static entropic energy'.
Defined by Eq.~(51) in
the global approach of Dutton (1973), it is also
the available energy examined in the production of a
local version of Dutton's theory by Pichler (1977). 
The function ${a}_{e2}$ is equivalent to
another form of `static entropic energy' described
by Livezey and Dutton (1976). Furthemore, it is also
the dry part of the `exergy' function suggested
by Karlsson (1990). Even if the exergy term is a generic
name used in modern thermodynamics to denote any of the
availability functions ${a}_{e1}$, ${a}_{e2}$
or $a_h$, depending on the system to be investigated,
the terms `available energy' or `available enthalpy' 
will be used in this paper.

The previous relationship between the thermodynamic
availability functions ${a}_{e1}$ or ${a}_{e2}$ and
their meteorological counterparts
has already been stated in M91,
where a possible application of a local available-enthalpy 
function $a_h$ to atmospheric energetics
was also considered, but where the local 
cycle was not derived.
It was mentioned that the
theory of $APE$ presented in Pearce (1978, 
hereafter  P78) was the
first application of the available-enthalpy concept to
atmospheric science. Pearce defined the global
$APE$ function such that $d/dt(APE)=d/dt(H - T_r S)$
applies to the whole atmosphere, with $T_r\approx251$~K.
Clearly, the global available enthalpy 
\begin{equation}
A_h \: = \: ( H - T_r \: S ) \: - \: ( H_r - T_r \: S_r)
\nonumber
\end{equation}
is a solution
to this equation, but the constant term $H_r - T_r S_r$
was missing in the paper of Pearce and his following
mathematical developments were carried out using several
approximations that will be overcome in this paper.

Other approaches have also been proposed in meteorology
to generalize the results of Lorenz, though most of
them have not succeeded in deriving a set of 
energy equations similar to the Lorenz cycle.
The same is also true for the
local and positive-definite potential 
energy of Andrews (1981) - see section 6
in Part-I of this paper for further explanations  --, for
the $APE$ of MacHall (1990), or for the 
local pseudo-energy concept of Shepherd (1993) 
-- a generalization of all meteorological availability 
functions -- also discussed in Kucharski (1997). 
As demonstrated by Marquet (1995), it is possible to
deduce Lorenz, Dutton or Pearce's results by
choosing, for each case, an appropriate reference
state in the pseudo-energy theory. However,
up to now, it appears that no general cycle 
has been published starting with this concept.

      \subsection{The available enthalpy function} 
      \label{subsection_2.3}

The basis for the research presented in this paper 
can be found in the thesis report of Marquet's (1994).
However, several theoretical improvements and some new
applications will be included in this article.
An approach similar to P78 will be retained by 
separating $A_h$ into three energy components,
depending on pressure averages for
$A_S$, zonally symmetric circulations for 
$A_Z$ and eddy circulations for $A_E$.
Kinetic energy will also be 
separated into three parts $K_S$, $K_Z$ and $K_E$, contrary
to L55 and P78 when $K_S$ and $K_Z$ were merged into
a single component, also called $K_Z$. The new proposal
is an available enthalpy cycle with $A3+K3$ components
($3$ for the thermal part and $3$ for the kinetic part), 
substituting $A2+K2$ in L55 and $A3+K2$ in P78.

 \section{\Large \underline{Energy components}.}
 \label{section_3}

      \subsection{Local available enthalpy $a_h$.} 
      \label{subsection_3.1}

According to M91, available enthalpy per unit mass `$a_h$' defined by (\ref{equa_ah}) is equal to $( h - h_r ) - T_r \: (s - s_r) $.
Therefore it only depends on differences in enthalpy and entropy which only depends on local temperature ($T$) and pressure ($p$):
\begin{align}
    h - h_r & = \;  c_p \:  ( T - T_r )
      \: ,  \label{defh} \\ 
    s - s_r & \; = \;  c_p \: 
           \ln \left\{ \left( \frac{T}{ T_r} \right)
                \left(\frac{p}{p_r} \right)^{-{ \kappa}} \right\}
         \; = \;
         c_p \: 
           \ln \left( \frac{T}{ T_r} \right)
           \: - \: 
           R  \:
           \ln \left(\frac{p}{p_r} \right)
      \: .  \label{defs}
\end{align}
Note that absolute values for enthalpy $h$ or even entropy 
$s$ need not be known, only the relative differences (\ref{defh}) 
and (\ref{defs}) are required for determining $a_h$.
The reference temperature and pressure $T_r$ and
$p_r$ are chosen as two constants in space and time.
Following M91, $1/T_r$ and $\ln(p_r)$ should
be global and long-range averages of
$1/T$ and $\ln(p)$, respectively. In fact, 
two prescribed numerical values, set to
$250$~K for $T_r$ and $1000/\exp(1) \approx 368$~hPa
for $p_r$, will be used in this paper.
It will be demonstrated later that 
results will not be affected when the
reference values are perturbed.

According to M91, the available enthalpy (\ref{equa_ah}) 
can be separated into a sum of two local 
components $a_T$ and $a_p$, the first one depending on temperature,
the other one on pressure, to give
\vspace{-0.15cm}
\begin{eqnarray}
  a_h \: (T, p \: ; \: T_r, p_r) 
      & = & 
      a_T \: (T \: ; \:  T_r) \: + \: a_p \:(p \: ; \: T_r, p_r)
      \: ,  \label{sepah1}
\end{eqnarray}
where
\begin{eqnarray}
  a_T \; = \;   
      c_p \; T_r \:{\cal F}( X ) \: ,
      \hspace{1.4cm} 
  a_p \; = \;
       R  \: \: T_r \:  \ln ( p / p_r ) 
      \: , \label{defaTap}
\end{eqnarray}
\begin{eqnarray}
  {\cal F}( X ) \: = \:  X - \ln (1+X) \: ,
       \hspace{0.4cm} 
  X ( T , T_r ) \; = \; \frac{ T - T_r }{ T_r}
                     \; = \; \frac{ T }{ T_r} - 1
      \: . \label{defFX}
\end{eqnarray}
The local temperature component $a_T$ is written with the help of
a function ${\cal F}$ defined by (\ref{defFX})
for any variable $X>-1$. 
Function ${\cal F}$ also verifies the
exact separating property (\ref{propFX1X2})
that holds whenever $ X_1 > -1$ and
$ X_2 > -1$ (in which case $ X_1 + X_2 +  X_1 \: X_2 = 
(1 + X_1 )  \: (1 + X_2) -1$ is also greater than $-1$):
\vspace{-0.15cm}
\begin{eqnarray}
       {\cal F} (X_1 + X_2 + X_1 \: X_2 ) \! \! & = & \! \! 
       {\cal F} (X_1) \; + \; {\cal F} (X_2) \; + \; X_1 \: X_2 \: .
               \label{propFX1X2}
\end{eqnarray}
Function ${\cal F}$ is a positive, quadratic function 
for small $|X|$, as indicated by the expansions (\ref{approxFXDVT})  
and (\ref{approxFAXDVT}). Typically, $|X| < 0.3$ for 
$ T_r  \equiv 250$~K and for temperatures
between $320$~K and $180$~K, as observed in usual atmospheric
conditions.
\vspace{-0.15cm}
\begin{eqnarray}
       {\cal F} (X) \! \! & = & \! \! 
       \frac{X^2}{2} \; - \; \frac{X^3}{3} \; + \; o(X^3) \: ,
               \label{approxFXDVT} \\
       \Rightarrow
       {\cal F}( X ) \! \! & \approx & \! \!
       {\cal G}( X )  \: = \: \frac{X^2}{2}
       \hspace{0.6cm} \hbox{for } \;
       |X| \; \approx 0 \: ,
               \label{approxFX} \\
       {\cal F} (\alpha \: X) \! \! & = & \! \! 
       {\alpha}^2 \: {\cal F} ( X ) \; - \; 
       {\alpha}^2 \: ( \alpha - 1) \: \frac{X^3}{3} \; + \;  o(X^3) \: .
               \label{approxFAXDVT}
\end{eqnarray}

As a consequence, ${\cal G}$ given by (\ref{approxFX}) 
is a good approximation of ${\cal F}$
for small $|X|$, and the temperature component 
(\ref{defaTap}) is found to be similar to the local function 
`$a$' in P78. The result is 
$a_T \approx a = c_p \: {(T-T_r)}^2 / (2 \: T_r)$.

       \subsection{Limited area components for $a_h$ and $e_k$} 
      \label{subsection_3.2}

According to L55 and the following limited-area 
applications of Muench (1965), Brennan and Vincent (1980),
P78 and Michaelides (1987), the eddy part of the flow 
will be computed by a departure from the zonal 
average circulation, when the zonal average
is defined over the limited area domain.
The notations for the isobaric $\overline{(...)}$ 
and zonal ${(...)}^{\lambda}$ averaging operators 
are described in Appendix-A, where 
superscripts and subscripts 
(for instance 
$T^{\lambda} = T - T_{\lambda}$) represent average 
values and deviations from them, respectively.

It is expected that $a_T$ can be separated 
into the three local components of P78, $a_S$, $a_Z$ and 
$a_E$, possibly with further local terms.
A concise description will be obtained in terms 
of the function ${\cal F}$ of $X_S$, $X_B$, $X_Z$ and $X_E$. 
The final result will be obtained with the
property (\ref{propFX1X2}) applied
successively to the exact separations
\vspace{-0.5mm}
\begin{eqnarray}
  \left( \frac{T }{ T_r} - 1 \right)
  \hspace{-0.2cm} \; & = & \; \hspace{-0.2cm}\; 
    \left( \frac{T  }{ \overline{T}}   - 1 \right) \; \; + \;
    \left( \frac{\overline{T}  }{ T_r} - 1 \right) \; \; + \;
    \left( \frac{T  }{ \overline{T}}   - 1 \right)
    \left( \frac{\overline{T}  }{ T_r} - 1 \right) \: , 
  \label{defsep1} \\ 
    \left( \frac{ T }{ \overline{T} }   - 1 \right) 
  \hspace{-0.2cm} \; & = & \; \hspace{-0.2cm}
    \left( \frac{ T }{ {T}^{\lambda} }   - 1 \right)  \; +\; 
    \left( \frac{ {T}^{\lambda} }{ \overline{T} }   - 1 \right) \; + \; 
   \left( \frac{ T }{ {T}^{\lambda}   }   - 1 \right) 
   \left( \frac{ {T}^{\lambda} }{ \overline{T} }   - 1 \right) \: .
  \label{defsep2}
\end{eqnarray}
Equations (\ref{defsep1}) and (\ref{defsep2})
can be understood as an insertion of $\overline{T}$
between $T$ and $T_r$ for (\ref{defsep1}), and an 
insertion of ${T}^{\lambda}$ between $T$ and 
$\overline{T}$ for (\ref{defsep2}). Note that these equations 
are directly put in the form $X_1 + X_2 + X_1 \: X_2$ as
required by (\ref{propFX1X2}).

After some manipulations, it is found that the 
temperature component $a_T$ can indeed be written as
a sum of the local version of Pearce components $a_S$, 
$a_Z$ and $a_E$, with two additional terms 
$a_{cS}$ and $a_{cZ}$, to give
\vspace{-0.5mm}
\begin{eqnarray}
  a_T  \! \! & = &  \! \! 
                \: a_S    \: + \: a_Z   \: + \: a_E\:
              + \: a_{cS} \: + \: a_{cZ} \: ,  
  \label{sepah2}
\end{eqnarray}
where
\vspace{-0.5mm}
\begin{align}
       &   
       a_S   \; = \; c_p \; T_r \; {\cal F}( X_S ) \: , \; \; \; \; \;
       a_Z   \; = \; c_p \; T_r \; {\cal F}( X_Z ) \: , \; \; \; \; \;
       a_E   \; = \; c_p \; T_r \; {\cal F}( X_E ) \: ,
  \label{defaSaZaE}
\end{align}
and
\begin{align}
       &  
       a_{cS} \; = \; c_p \; T_r \;  X_S \; X_B  
       \: ,   \; \;\;  
       X_S    \; = \; \frac{{\overline{T} - T_r}}{T_r}
       \: ,   \; \;\;  
       X_B    \; = \; \frac{T - \overline{T}}{\overline{T}}
       \: ,
  \label{defacS} \\ 
       & 
       a_{cZ} \; = \; c_p \; T_r \;  X_Z \; X_E        
       \: ,   \; \;\;  
       X_Z    \; = \; \frac{{T}^{\lambda} - \overline{T}}{\overline{T}}
       \: ,   \; \;\; 
       X_E    \; = \; \frac{ T - {T}^{\lambda}}{{T}^{\lambda}} \: ,
  \label{defacZ} 
\end{align}
or, alternatively,
\begin{eqnarray}
       & & \hspace{-0.5cm}     
       X_B    \; = \; \frac{T^{\:\prime}}{\overline{T}}
       \: ,   \; \;\;  \hspace{1.1cm}  
       X_Z    \; = \; \frac{{T}^{\lambda}_{\varphi}}{\overline{T}}
       \: ,   \; \;\;  \hspace{1.1cm}  
       X_E    \; = \; \frac{{T}_{\lambda}}{{T}^{\lambda}} \: .
  \label{defaaux}
\end{eqnarray}

The function ${\cal F}(X)$ is always positive and
equal to zero only if $X = 0$. This property can
be applied to the mean values $\overline{a_S}$, $\overline{a_Z}$ 
and $\overline{a_E}$ which differ from zero
only if $\overline{T} \neq T_r$, ${T}^{\lambda} \neq \overline{T}$ and
$T \neq {T}^{\lambda}$, respectively.

The components of Pearce are obtained as approximate forms
of Eqs.~(\ref{defaSaZaE}) when ${\cal F}$ 
is replaced by ${\cal G}$, together with the hypotheses 
${(T_r/\overline{T})}^2 \approx 1$ and 
${(T_r/{T}^{\lambda})}^2 \approx 1$, giving
\begin{eqnarray}
 & & \hspace{-1.0cm} 
 a_S \; \approx \; 
 c_p \; \frac{ \left( \overline{T} - T_r \right)^2 }{2 \: T_r} \; , 
 \hspace{0.5cm}
 a_Z \; \approx \; 
 c_p \; \frac{ \left( {T}^{\lambda}_{\varphi} \right)^2  }{ 2 \: T_r}  \; ,
 \hspace{0.5cm}
 a_E \; \approx \; 
  c_p \; \frac{  \left( {T}_{\lambda} \right)^2  }{ 2 \: T_r}  \: .
 \label{defappP78}
\end{eqnarray}

The three components $a_S$, $a_Z$ and $a_E$ have
been called in P78, `static stability', `zonal' and
`eddy' reservoirs, respectively.
The new component $a_p$ given by (\ref{defaTap}) and
the two complementary parts $a_{cS}$ and $a_{cZ}$ in
(\ref{sepah2}) were missing in P78. There was no impact on
the global scale since the vertical integral of $a_p$ 
and the horizontal average $\overline{a_{cS}}$ 
and $\overline{a_{cZ}}$ 
are $0$. But, for a limited area study, the flux of these 
additional components $\overline{B(a_p)}$,
$\overline{B(a_{cS})}$ and $\overline{B(a_{cZ})}$ 
are non zero and cannot be neglected.

It is also possible to separate the kinetic energy 
$e_k={\bf U}_h . {\bf U}_h /2$ into
three components $k_S$, $k_Z$ and $k_E$, with two 
additional parts $k_{cS}$ and $k_{cZ}$, giving 
\begin{align}
  e_k    & = \; 
         k_S \; + k_Z \; + \; k_E \; + \; k_{cS}\; + \; k_{cZ}
      \: ,   \label{sepek}
\end{align}
where
\begin{align}
  k_S    & = \;
      \frac{ (\overline{u}){}^2 + (\overline{v}){}^2 }{  2 }   
  \: , \hspace*{5mm}
  k_Z    \; = \; 
      \frac{ ({u}^{\lambda}_{\varphi}){}^2 
             + ({v}^{\lambda}_{\varphi}){}^2 }{  2 } 
  \: ,  \hspace*{5mm}
  k_E    \;  = \; 
      \frac{  ({u}_{\lambda}){}^2 + ({v}_{\lambda}){}^2 }{  2 }    
      \: ,   \label{def_KSZE} \\   
  k_{cS} & = \; 
      u^{\prime} \: \overline{u} \; + \; v^{\prime} \: \overline{v} \:  
      \: ,   \hspace*{4mm}
  k_{cZ} \;  = \; 
         \: {u}^{\lambda}_{\varphi} \; {u}_{\lambda} 
    \; + \; {v}^{\lambda}_{\varphi} \; {v}_{\lambda} \:       
         \: .  \label{defkC}
\end{align}
An exact separating property (\ref{propGX1X2}) observed for the quadratic function ${\cal G}(X) = X^2/2$, equivalent to (\ref{propFX1X2}) observed for ${\cal F}(X) = X - \ln(1+X)$ and $a_h$, has been used to derive  (\ref{def_KSZE})-(\ref{defkC}):
\begin{eqnarray}
       {\cal G} (X_1 + X_2) \! \! & = & \! \! 
       {\cal G} (X_1) \; + \; {\cal G} (X_2) \; + \; X_1 \: X_2
   \label{propGX1X2}  \: .
\end{eqnarray}

        \subsection{Interpretations for the limited-area components.} 
      \label{subsection_3.3}

The six components ($a_S$, $a_Z$, $a_E$) and
($k_S$, $k_Z$, $k_E$) are little known in atmospheric
energetics. Even if the three available-enthalpy reservoirs
are similar to the components defined in P78,
results published in the global
approach of Pearce have not been widely applied
in meteorology and it is worthwhile to explore further
their physical meaning, especially
for a limited-area domain. The same is true for the separation
of $e_k$ into three components where the eddy part, $k_E$,
is the only part not to undergo a redefinition ; it
is defined as usual. The large-scale parts $k_Z$ and
$k_S$ correspond to a new approach justified for the sake
of retaining symmetry between the two sets of components.

\begin{figure}[hbt]
\centering
(a) \hspace*{2mm} $\overline{T}-T_r$ for $a_S$
\hspace*{4 cm} 
(b) \hspace*{2mm} ${T}({\lambda}, {\varphi})$
\hspace*{2 cm} 
\\
\includegraphics[width=0.33\linewidth,angle=0,clip=true]{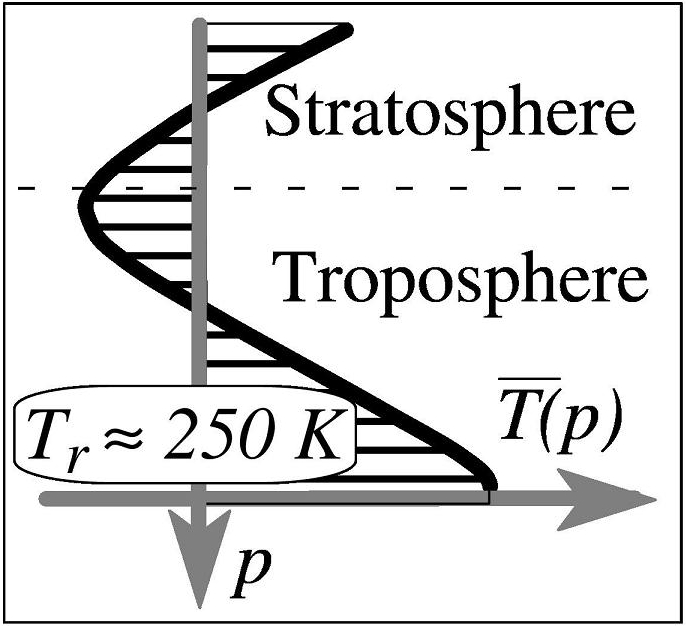}
\includegraphics[width=0.50\linewidth,angle=0,clip=true]{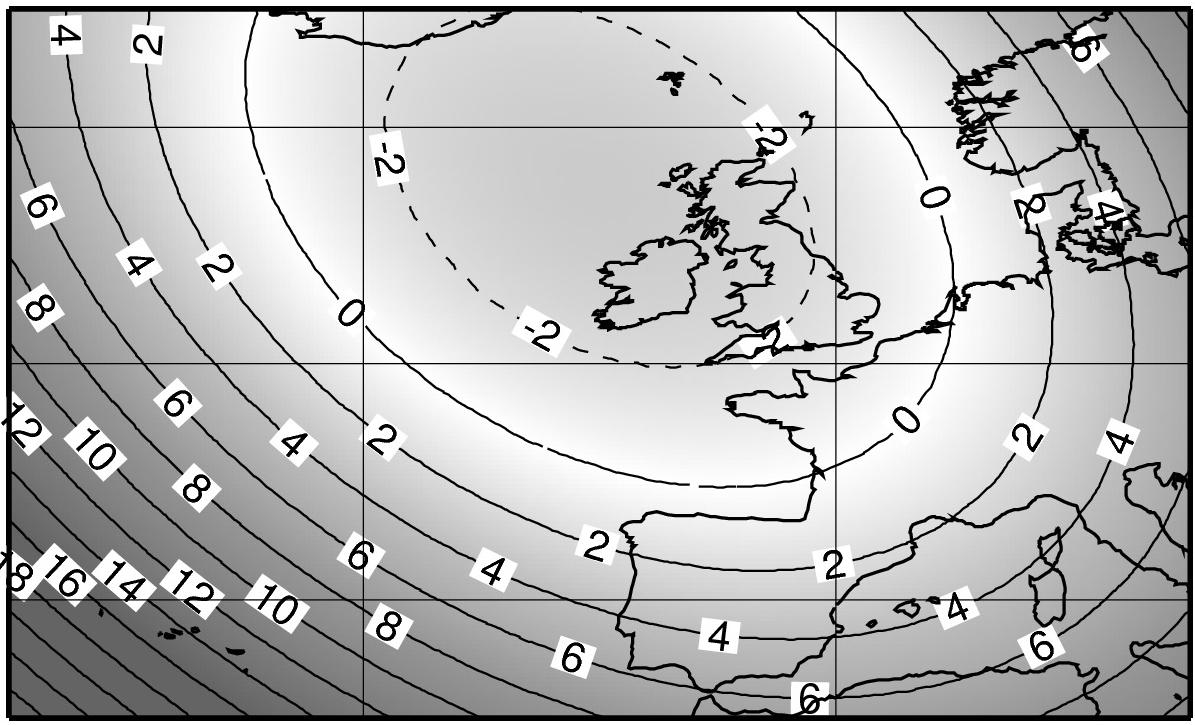}
\\
\vspace{2mm}
(c) \hspace*{2mm} ${T}^{\lambda}_{\varphi}$ for $a_Z$
\hspace*{5 cm} 
(d) \hspace*{2mm} ${T}_{\lambda}$ for $a_E$
\\
\includegraphics[width=0.48\linewidth,angle=0,clip=true]{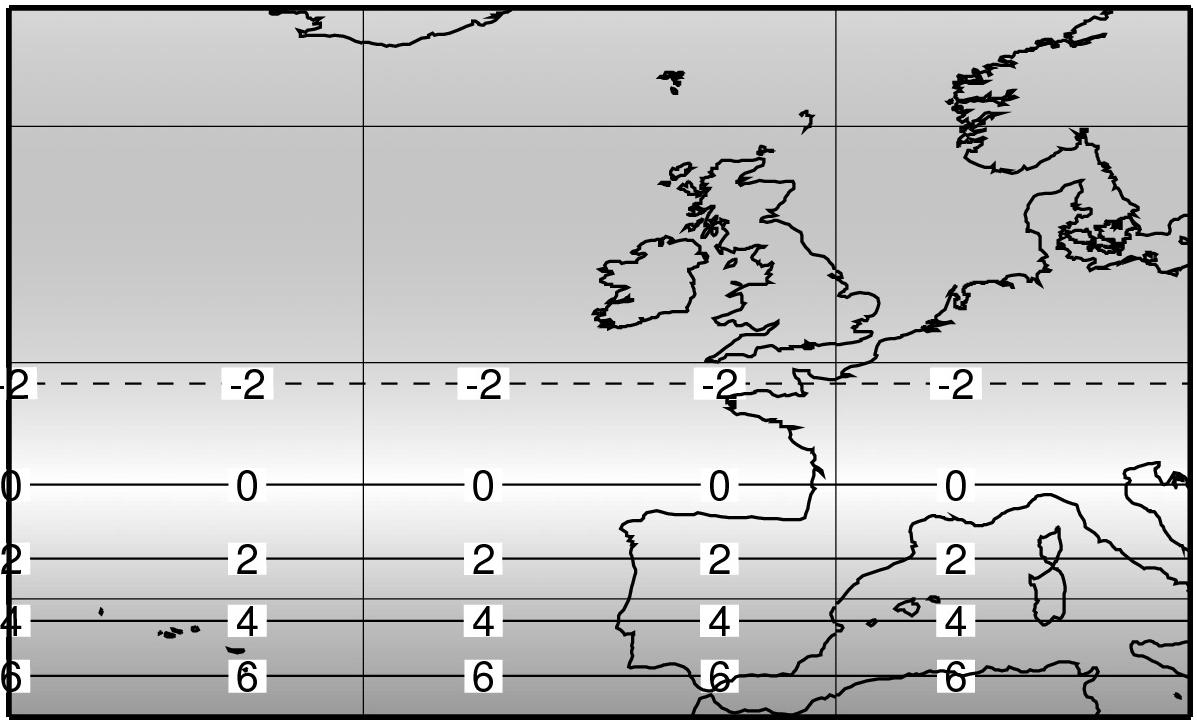}
\includegraphics[width=0.48\linewidth,angle=0,clip=true]{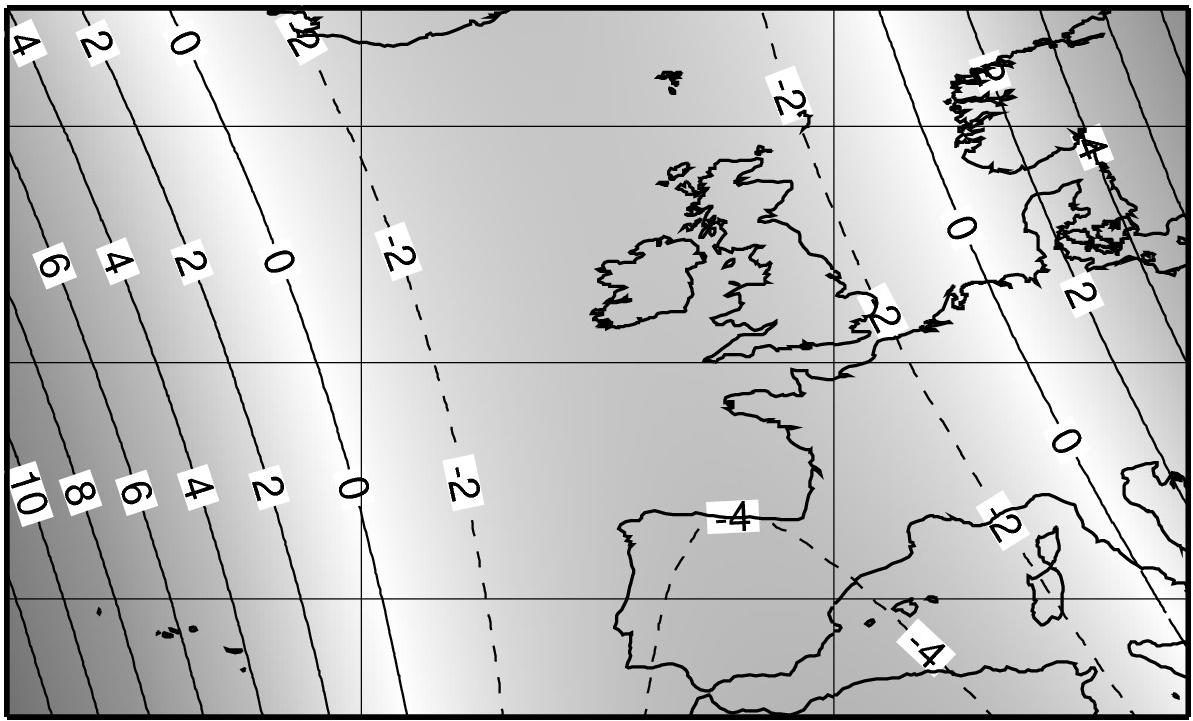}
\vspace{-2mm}
\caption{\it \small
The separation of $\overline{a_T}$ (the thermal part of the
available enthalpy $\overline{a_h}$ ) into $\overline{a_S} \: + \:
\overline{a_Z} \: + \: \overline{a_E}$. 
(a)~Values of $\overline{T}-T_r$ for $a_S$.
(b) A temperature distribution made of a 
cold minimum located within a limited area domain.
(c) Values of ${T}^{\lambda}_{\varphi}$ for $a_Z$.
(d) Values of ${T}_{\lambda}$ for $a_E$.
See text for further explanation.
\label{FigSEPAH}}
\end{figure}

An example of $\overline{a_T}$ separation is shown on
Fig.~\ref{FigSEPAH}. Following Pearce, `$a_S$' will be called
the vertical `static stability' component and according
to Fig.~\ref{FigSEPAH}(a) is the part of 
thermal availability created by the
difference between the average vertical profile
$\overline{T}$ and a constant prescribed profile $T_r$
(shaded area).
The horizontal separation of $\overline{a_T}$ into 
$\overline{a_Z} + \overline{a_E}$
is illustrated by the use of a simulated temperature
distribution, described in Fig.~\ref{FigSEPAH}(b).
It is an elongated cold minimum with a north-west to
south-east orientation and 
the minimum is not centred
with respect
to the limited area domain. 
Figure~\ref{FigSEPAH}(c) shows how 
the component $a_Z$ is 
created by the north/south differences 
in zonal average temperature
${T}^{\lambda}_{\varphi}={T}^{\lambda}-\overline{T}$. The
isopleths in Fig.~\ref{FigSEPAH}(d) represent the distribution
of the zonal departure ${T}_{\lambda} = T - {T}^{\lambda}$
which generates the eddy component $a_E$. Even if the
east/west gradient prevails, it is found that the north-west 
to south-east tilted feature is still present.

The old Lorenz partitioning  $APE = A_Z + A_E$
corresponds to a mixing of horizontal and 
vertical departure terms defined by the integrands
${T}^{\lambda}_{\varphi}/\overline{\sigma}$ and 
${T}_{\lambda}/\overline{\sigma}$, respectively,
where $\overline{\sigma}$ is the mean static stability.
The numerators for these integrands are 
represented on Fig.~\ref{FigSEPAH}(c)
and (d) for ${T}^{\lambda}_{\varphi}$ and
${T}_{\lambda}$, respectively.
The common denominator $\overline{\sigma}$
depends on the mean vertical lapse rate and
can be compared to some extent with the
component $a_S$ which depends on $\overline{T} - T_r$. 
However an important result is that the
component $a_S$ is still valid for hydrostatically
neutral or unstable states when local values of
$\sigma$ are close to zero or negative, in which
case local values of Lorenz's components 
depending on $\overline{\sigma}$ become infinite and meaningless.

\begin{figure}[thb]
\centering
(a) \hspace*{2mm} Wind ($u$, $v$) for $e_k$
\\
\includegraphics[width=0.40\linewidth,angle=0,clip=true]{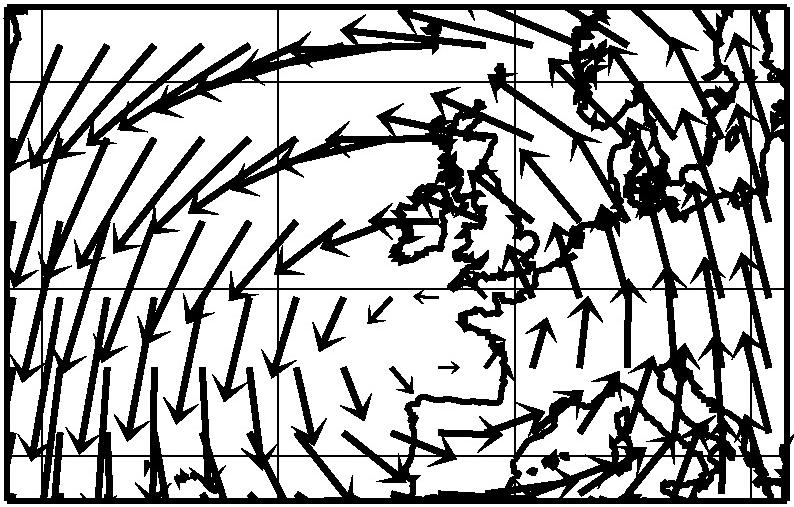}
\\
(b) \hspace*{2mm} (${u}_{\lambda}$, ${v}_{\lambda}$) for $k_E$
\hspace*{45mm} 
(c) \hspace*{2mm} (${u}^{\lambda}_{\varphi}$, ${v}^{\lambda}_{\varphi}$) 
                   for $k_Z$
\\
\includegraphics[width=0.40\linewidth,angle=0,clip=true]{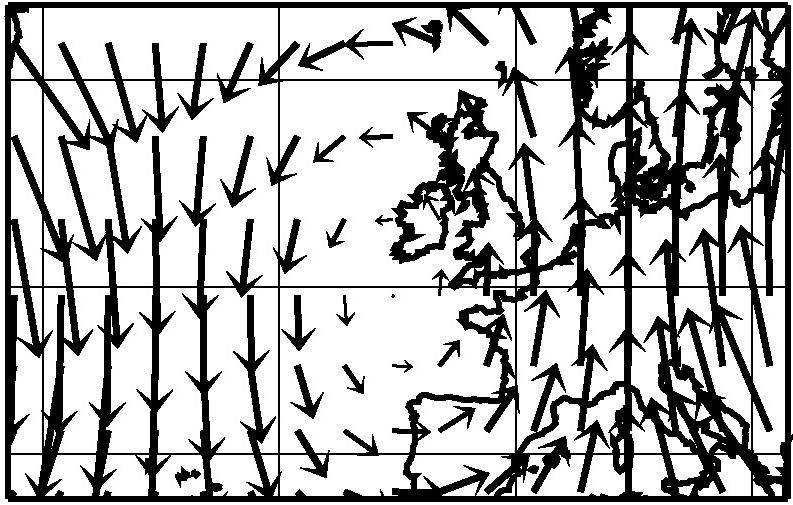}
\hspace*{20mm}
\includegraphics[width=0.40\linewidth,angle=0,clip=true]{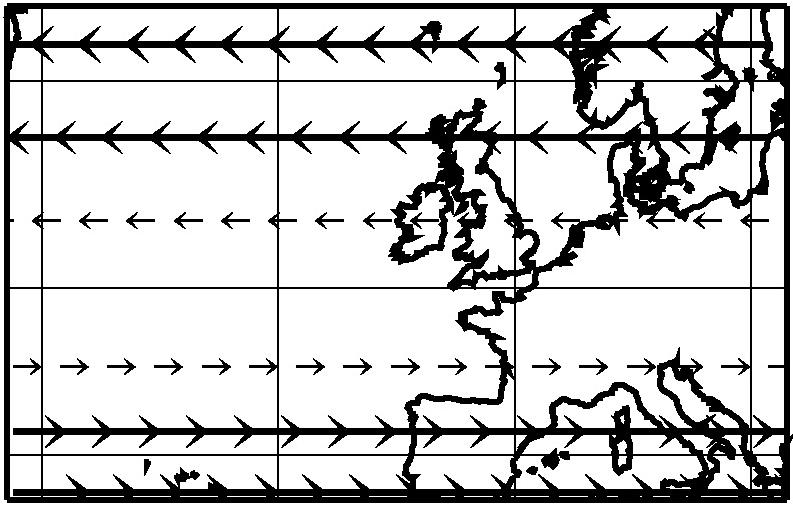}
\\
\hspace*{2 cm} 
(d) \hspace*{2mm} ($\overline{u}$, $\overline{v}$) for $k_S$
\hspace*{30mm} 
(e) \hspace*{2mm} Lorenz and Pearce `${K_Z}$'$\: = k_S+k_Z$
\\
\includegraphics[width=0.40\linewidth,angle=0,clip=true]{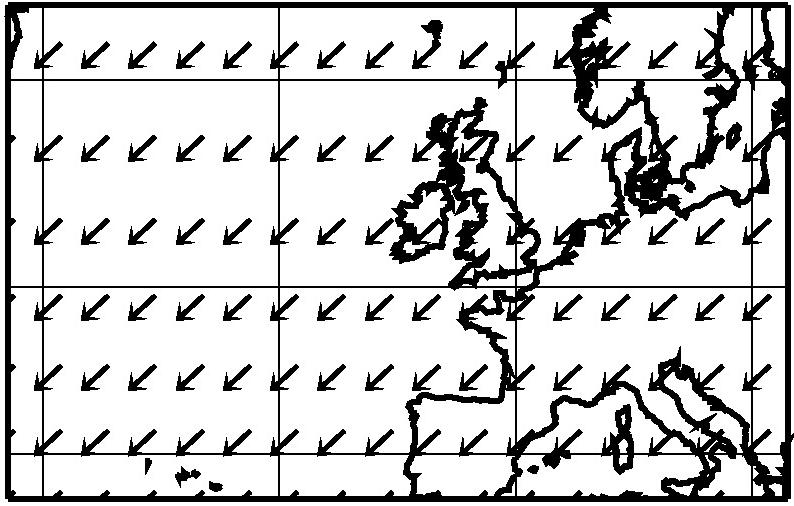}
\hspace*{20mm}
\includegraphics[width=0.40\linewidth,angle=0,clip=true]{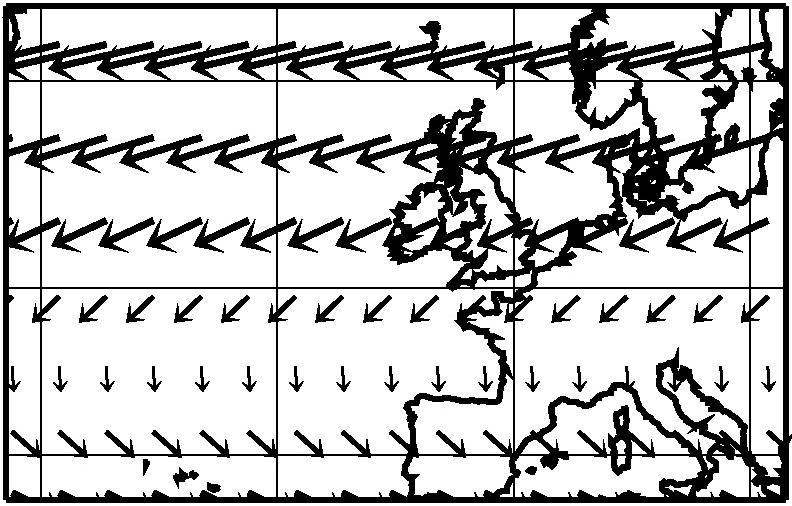}
\vspace{-2mm}
\caption{\it \small
An illustration for the separation (\ref{sepek})
applied to a simulated vortex within
a limited area domain. The separation
$\overline{e_k} = \overline{k_S} + \overline{k_Z} 
+ \overline{k_E}$ proposed in 
this paper is shown in (b), (c) and (d). Lorenz's 
and Pearce's large scale version $\overline{k_S} + \overline{k_Z} $ 
is depicted in (e).
(a) The map for the
local wind ($u$, $v$).
(b) The wind (${u}_{\lambda}$, ${v}_{\lambda}$) for $ k_E$. 
(c) The wind (${u}^{\lambda}_{\varphi}$, ${v}^{\lambda}_{\varphi}$)
for $k_Z$. 
(d) The wind ($\overline{u}$, $\overline{v}$) for $k_S$. 
(e) Lorenz and Pearce large scale component
for the wind (${u}^{\lambda}$, ${v}^{\lambda}$)
for $k_S + k_Z = \:$`${K_Z}$'.
See text for further explanation.
\label{FigSEPKIN}}
\end{figure}

Figure~\ref{FigSEPKIN} shows the new separation 
of $\overline{e_k}$ into 
$\overline{k_E} + \overline{k_Z} + \overline{k_S}$
when it is applied to a simulated vortex not centred 
with respect to the limited-area domain. The local wind 
vectors corresponding to components $k_E$, $k_Z$ and $k_E$ are 
depicted by Figure~\ref{FigSEPKIN}~(b), 
(c) and (d), 
respectively. The eddy-vortex part 
of the flow is clearly 
captured by Figure~\ref{FigSEPKIN}~(b)
(${u}_{\lambda}$ and ${v}_{\lambda}$). Furthermore,
large scale shearing of the zonal average wind
can be recognized on Figure~\ref{FigSEPKIN}~(c)
(${u}^{\lambda}_{\varphi}$ and ${v}^{\lambda}_{\varphi}$).
The third component $k_S$ corresponds to
the uniform average motion $\overline{u}, \overline{v}$.
The wind field representation, $k_Z + k_S$ (depicted 
on Fig.~\ref{FigSEPKIN}(e)), is the component 
referred to as `$K_Z$' by Lorenz and Pearce. For
this particular flow in this limited area,
a large scale vortex feature is visible for `$K_Z$' and
it is redundant with the $k_E$ information.
Moreover, the average motion picture shown on
Fig.~\ref{FigSEPKIN}(d) is not easily identified in
Fig.~\ref{FigSEPKIN}(e). For these reasons,
the new separation into the
three components $\overline{k_E} + \overline{k_Z} + 
\overline{k_S}$ as proposed in this study seems to be 
appropriate to catch relevant spatial scales 
for such a vortex motion.

 \section{\Large \underline{The limited-area available enthalpy cycle}.}
 \label{section_4}

      \subsection{Basic equations.} 
      \label{subsection_4.1}

The available-enthalpy cycle will be reproduced as
a set of six equations for the six 
available-enthalpy and kinetic-energy components.
Pressure coordinates will be used with vertical velocity 
$\omega = d/dt(p)$ according to Kasahara (1974).
The Eulerian time derivative operator
$\partial / \partial t$ will be applied to
each of the six components and will be
expressed by using the material derivative
$d/dt$ and a boundary function $B$. 
The resulting operator is given for any 
scalar $\eta$ as
\begin{eqnarray}
      \frac{\partial \, \eta}{\partial \, t}
   & = & 
      \frac{d \,  \eta}{d \, t}
      \: - \: B(\eta)
      \: , \label{defderiv}
\end{eqnarray}
where
\vspace{-0.15cm}
\begin{eqnarray}
   B(\eta) 
      \hspace{-0.2cm} & = & \hspace{-0.2cm} 
      {{\rm div} \,}_{\!\!p} \: ( \eta \: {\bf U}_h \: )  
      \; + \; \frac{\partial}{\partial \, p}
                \left( \eta \: \omega \right)
   \; = \;        {\bf U}_h \: . \: {\bf \nabla}_{\!p} \:(\eta)
      \; + \; \omega \; \frac{\partial \, \eta}{\partial \, p}
      \: , \label{defBeta} \\
   \mbox{with} \hspace{1.0cm} B(\phi) 
      \hspace{-0.2cm} & = & \hspace{-0.2cm} 
      {\bf U}_h \: . \: {\bf \nabla}_{\!p} \:(\phi)
      \; - \; \frac{R}{p} \; \omega \: T
      \: . \label{defBPhi}
\end{eqnarray}
The hydrostatic assumption and
continuity equations will be used in the forms:
\vspace{-0.15cm}
\begin{eqnarray}
      \frac{\partial \, \phi}{\partial \, p}
      \; = \; - \; \frac{R\: T}{p}
\hspace{0.7cm} \mbox{and} \hspace{0.7cm}
      0 \; = \;  {{\rm div} \,}_{\!p} \: ( {\bf U}_h \: ) 
      \; + \; \frac{\partial \, \omega}{\partial \, p}
\: . \label{defhydroconti}        
\end{eqnarray}
The two representations of $B(\eta)$ in Eq.~(\ref{defBeta})
in terms of divergence or gradient operators are
equivalent, linked by the continuity equation. 
The special case for 
$B(\Phi)$ in Eq.~(\ref{defBPhi})
is obtained when the hydrostatic assumption is taken 
into account.

The momentum and thermodynamic equations as used in
the Eulerian version of the French 
Arpege\footnote{\color{blue} ``Action de Recherche 
Petite Echelle Grande Echelle''.
It is the French counterpart of the ECMWF-IFS model.}
model can be 
written as follows (Courtier {\em et al.}, 1991) :
\vspace{-0.15cm}
\begin{eqnarray}
    \frac{d \, {\bf U}_h}{d\, t} \! & = & \!
               - \: {\bf \nabla}_{\!p}(\phi)
               \; - \; f \: {\bf k} \times {\bf U}_h
               \; + \; {\bf F}_h
   \: , \label{defevolUh}  \\    
    \frac{d \, u}{d\, t} \!  & = & \!
                  - \: [{\bf \nabla}_{\!p}(\phi)]{}_{x}
               \; + \; f^{\ast} v  
               \; + \; ({F}_h){}_x
   \: , \label{defevolUu}  \\    
    \frac{d \, v}{d\, t} \!  & = &  \!
                  - \: [{\bf \nabla}_{\!p}(\phi)]{}_{y}
               \; - \; f^{\ast} u  
               \; + \; ({F}_h){}_y
   \: , \label{defevolUv}  \\    
   c_p \; \frac{d \, T}{d\, t} \!  & = & \!
               \; \frac{R}{p} \; \omega \: T
               \; + \; \dot{q}
   \: , \label{defevolT}    
\end{eqnarray}
where friction (${\bf F}_h$) and diabatic heating 
($\dot{q}$) are forcing terms.
The pseudo-Coriolis factor $f^{\ast}$ is the
sum of the usual Coriolis
($f = 2 \: \Omega \sin (\varphi)$) term
and horizontal curvature, giving
$ f^{\ast} = f + u \tan (\varphi) / {\cal R}$.

       \subsection{The energy equations.} 
      \label{subsection_4.2}

The kinetic energy equation per unit mass
is easily obtained for 
\begin{equation}
e_k \; = \; \frac{1}{2} \; {\bf U}_h \, . \, {\bf U}_h 
    \; = \; \frac{1}{2} \; \left( u^2+v^2 \right)
\nonumber
\end{equation}
by taking the dot product of Eq.~(\ref{defevolUh}) 
by ${\bf U}_h$, to get
\vspace{-0.15cm}
\begin{eqnarray}
   \frac{d \, e_k}{d\, t}
    & = & 
   - \: {\bf U}_h \: . \: {\bf \nabla}_{\!p} \:(\phi)
   \; + \; {\bf U}_h \: . \: {\bf F}_h
   \; = \; - \: B(\phi) \; - \; \frac{R}{p} \; \omega \: T
        \: - \;d
   \: . \label{defevolek}            
\end{eqnarray}
The frictional dissipation $d$ denotes
the scalar product $- {\bf U}_h \: . \: {\bf F}_h $
and the Coriolis term does not contribute to
any local exchange of energy. The second formulation
for (\ref{defevolek}) is a consequence of
Eq.~(\ref{defBPhi}) by which
$- \: {\bf U}_h \: . \: {\bf \nabla}_{\!p} \:(\phi)$
can be transformed into $- \: B(\phi) - R \: \omega \: T / {p}$.

The potential-energy equation per unit mass satisfies
\begin{eqnarray}
    \frac{d \,\phi}{d\, t}
     & = & 
             \frac{\partial \, \phi}{\partial \, t}
             \: + \:  B(\phi) 
             \: . \label{defevolPhi}    
\end{eqnarray}

The entropy equation is deduced from 
$s = s_{00} + c_p \ln \{ (T/T_{00}) {(p/p_{00})}^{-R/c_p } \}$
and used with Eq.~(\ref{defevolT}), to give:
\vspace{-0.15cm}
\begin{eqnarray}
   T \; \frac{d \,s}{d\, t}
    & = &
    c_p \; \frac{d \,T}{d\, t}
    \; - \; \frac{R}{p} \; \omega \: T
    \; \; \; = \; \; \;  \dot{q}
    \: . \label{defevols}     
\end{eqnarray}

An equation for $a_h$ is then obtained by applying the
Eulerian time derivative operator to (\ref{equa_ah}) 
with the use of (\ref{defevolT}) and (\ref{defevols}),
whilst remembering the fact that
the material derivatives of constant
terms $T_r$ and $p_r$ cancel out.
The result is
\vspace{-0.15cm}
\begin{eqnarray}
    \frac{d \,a_h}{d\, t}
    & = & 
             c_p \left( 1 - \frac{T_r}{T} \right) \: \frac{d \,T}{d\, t}
             \: + \: \frac{R}{p} \; \omega \: T_r
    \; = \;  \frac{R}{p} \; \omega \: T
             \: + \:  \left( 1 - \frac{T_r}{T} \right) \:   \dot{q}
             \: . \label{defevolah}    
\end{eqnarray}

The final term in (\ref{defevolah}) is a generation 
of available enthalpy by diabatic heating $\dot{q}$ with a modulation 
by the local efficiency factor $(1 - T_r/T)$, 
also called the Carnot factor in thermodynamics.
The sign of this factor is the same as that of
($T - T_r$), but the sign of the complete term 
$(1 - T_r/T) \: \dot{q}$ depends on
the correlation between $(1 - T_r/T)$ and $\dot{q}$.
The last term in (\ref{defevolah}) 
is interpreted as a generation by horizontal and 
vertical differential heating, as in L55 and P78.

The available enthalpy equation (\ref{defevolah})
is associated with a local law of conservation, valid
along any streamline.
The change in time of the sum $a_k + e_k + \phi$ is 
evaluated from (\ref{defevolek}), (\ref{defevolPhi})
and (\ref{defevolah}), to give
\vspace{-0.15cm}
\begin{eqnarray}
    \frac{d }{d\, t} \left( \: a_h + e_k + \phi \: \right)
    & = & 
       \frac{\partial \, \phi}{\partial \, t}
       \: + \:  \left( 1 - \frac{T_r}{T} \right)  \: \dot{q}
       \; - \;  d
       \: . \label{defBernoulli}    
\end{eqnarray}
As a result, the sum ($\: a_k+e_k+\phi \:$) is a constant 
along any particular streamline
for a frictionless and isentropic 
steady flow. It is Bernoulli's law,
valid for the available enthalpy.
The only difference from the usual Bernoulli's
equation observed for the sum $(\: h + e_k + \phi \:)$ is the Carnot's Factor $(1 - {T_r}{/T})$ in factor 
of the heating rate $\dot{q}$.

       \subsection{The limited-area available-enthalpy cycle.} 
      \label{subsection_4.3}

The budget equations for the new 
available-enthalpy cycle are obtained by computing the 
time derivatives of the six components
$a_S$, $a_Z$, $a_E$, $k_S$, $k_Z$ and $k_E$.
This requires considerable manipulations based on 
the definitions given in the previous sections and
in the Appendix-A.\footnote{\color{blue}
This result is already described in my PhD thesis Marquet (1994).
The large gap of several years between my PhD thesis 
and this QJRMS paper is due to discouraging comments 
from referees and others, and to a change in position
to join the Climate research team at CNRM.
The Prud'homme prize received in (1995) from the 
French Meteorological Society, unpublished results 
obtained during FASTEX, and possible applications of 
the available-enthalpy cycle to Climate Change were 
encouraging enough to make me submit these results.
}
There is no approximation, no
development in series and no missing terms.
An example of the beginning of the $a_Z$
computations is presented in Appendix-B.
All terms are rearranged to reproduce
the form of classical results for the main 
global-scale conversion, generation and 
dissipation terms.\footnote{\color{blue} 
Doubts are often expressed about the possibility 
and the relevancy of these rearrangements.
These form classical issues of atmospheric 
energetics expressed in terms of 
(closed or open) ``energy cycles''.
Moreover, energy reservoirs associated with exergy and
available enthalpy may be viewed as ``fictitious''.
It was the word used in an internal report of
students of the French School of Meteorology
directed by Jean-Philippe Lafore and Jean-Luc Redelsperger
(F. Engel, B. Petit and M. Pontaud, 1992).
Differently, I consider that in spite of difficulties for
interpreting some terms, this $a_h$-cycle derived and
motivated by these criticisms expressed in 1992
is relevant, simply because
i) classical results obtained by Lorenz and Pearce
are included in this available-enthalpy cycle, and
ii) all other terms are expressed as divergence of 
fluxes which mainly vanish in global applications.
} 
The final result
is as follows
\begin{equation}
\left.
\begin{aligned}
    \overline{ {\partial}_t ( a_S )  }  
       \; & \: = \;
            \; - \; \overline{  B( a_S + a_{cS} ) } 
            \; - \; \overline{  c_{AS} } 
            \; - \; \overline{  c_S  }        \mbox{\hspace{1.2 cm}}
            \; - \; \overline{  B(a_{p}) }    \mbox{\hspace{0.1 cm}}
            \; + \; \overline{g_S} \vspace*{-0.4 cm} \\
    \overline{ {\partial}_t ( a_Z ) } 
       \; & \: = \;
            \; - \; \overline{  B( a_Z + a_{cZ} ) } 
            \; + \; \overline{  c_{AS}  }
            \; - \; \overline{  c_Z  }
            \; - \; \overline{  c_A  }        \mbox{\hspace{1.7 cm}}
            \; + \; \overline{g_Z} \vspace*{-0.4 cm} \\
    \overline{ {\partial}_t ( a_E ) } 
       \; & \: = \;
            \; - \; \overline{  B( a_{E} ) }  \mbox{\hspace{2.3 cm}}
            \; - \; \overline{  c_E  }        \mbox{\hspace{0.1 cm}}
            \; + \; \overline{  c_A  }        \mbox{\hspace{1.6 cm}}
            \; + \; \overline{g_E} \vspace*{-0.4 cm} \\
    \overline{ {\partial}_t ( k_S ) } 
       \; & \: = \;
            \; - \; \overline{ B( k_{S} + k_{cS} )} 
            \; - \; \overline{ c_{KS} }       \mbox{\hspace{0.05 cm}}
            \; + \; \overline{ c_S  }         \mbox{\hspace{1.2 cm}}
            \; - \; \overline{{B( \phi )}_S}
            \; - \; \overline{d_S} \vspace*{-0.4 cm} \\
    \overline{ {\partial}_t ( k_Z ) }  
       \; & \: = \;
            \; - \; \overline{ B( k_Z + k_{cZ} )} 
            \; + \; \overline{ c_{KS} }
            \; + \; \overline{ c_Z }
            \; - \; \overline{ c_K }
            \; - \; \overline{{B( \phi )}_Z}
            \; - \; \overline{d_Z} \vspace*{-0.4 cm} \\
    \overline{ {\partial}_t ( k_E ) } 
       \; & \: = \;
            \; - \; \overline{ B( k_{E} )}   \mbox{\hspace{2.35 cm}}
            \; + \; \overline{ c_E }          
            \; + \; \overline{ c_K }           
            \; - \; \overline{{B( \phi )}_E}
            \; - \; \overline{d_E}
\end{aligned}
\;\;
\right\} 
\label{eq:cycle2}
\end{equation}

\begin{figure}[hbt]
\centering
(a) \hspace*{2mm}  Pearce's cycle (1978)
\hspace*{3 cm} 
(b) \hspace*{2mm} $AH$-cycle (Marquet, 1994)
\vspace*{1mm}
\\
\includegraphics[width=0.49\linewidth,angle=0,clip=true]{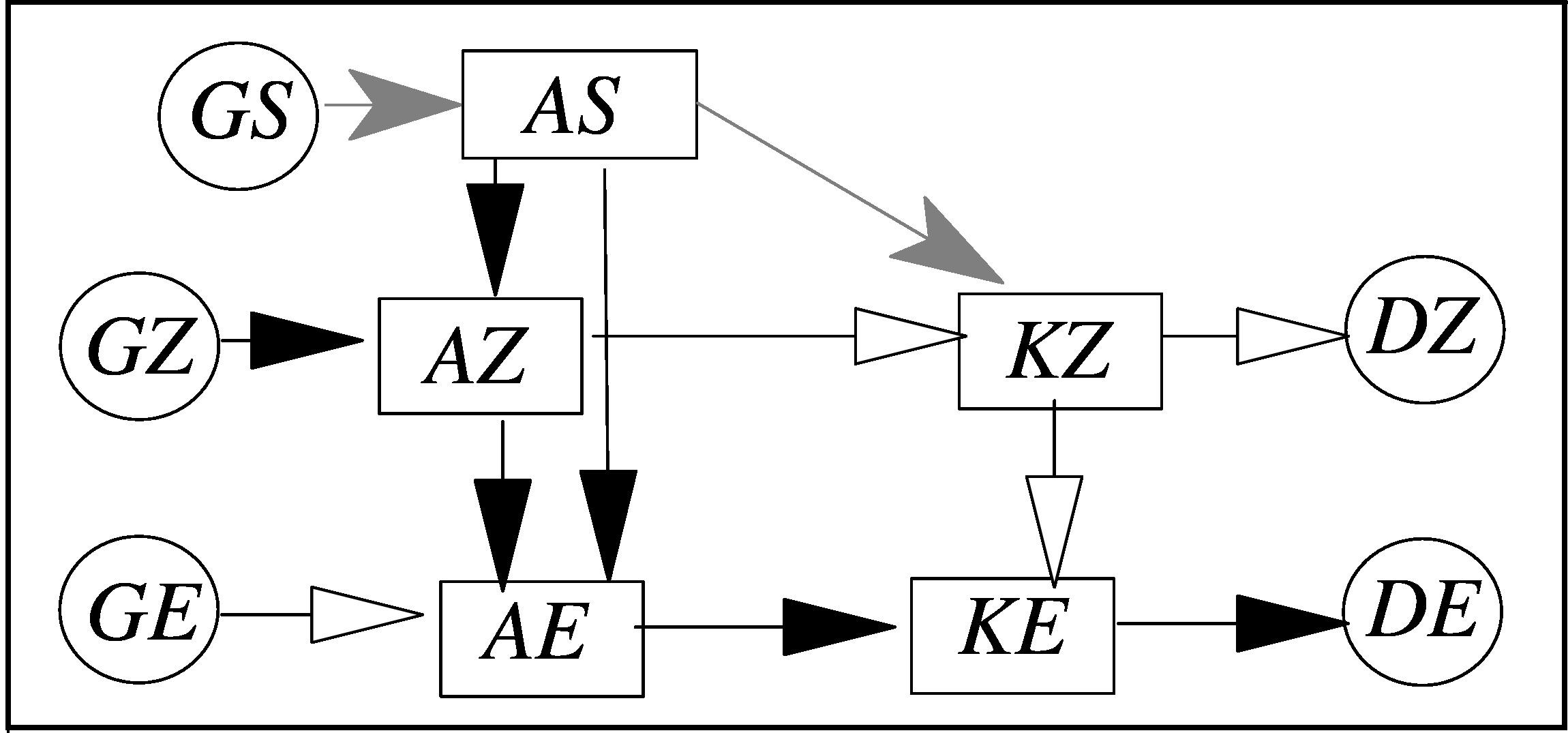}
\includegraphics[width=0.49\linewidth,angle=0,clip=true]{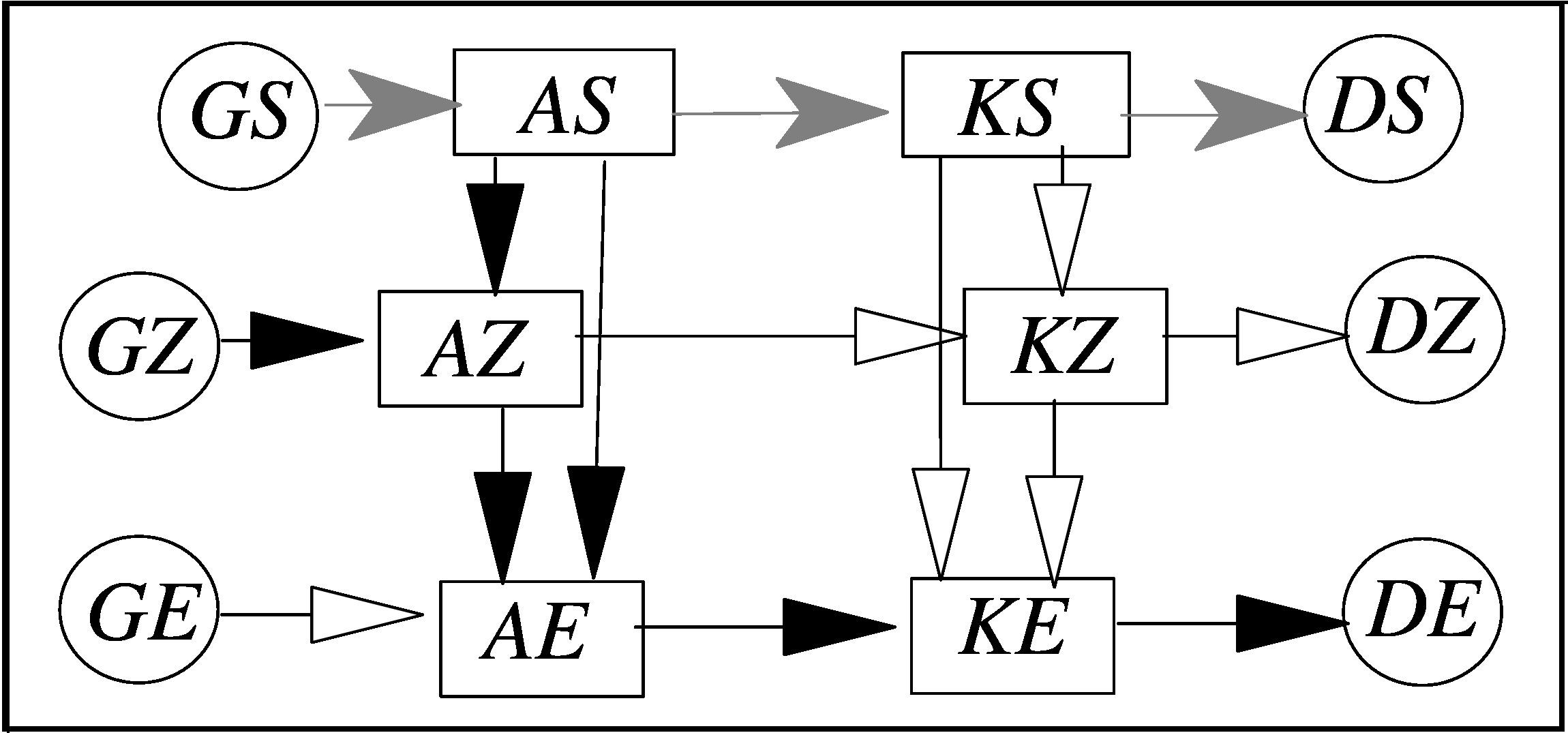}
\vspace*{-2mm}
\caption{\it \small
(a) Global and asymmetric $APE$ cycle diagram according to
Pearce (1978) ; 
(b) A possible version for a symmetric 
and global $AH$ cycle diagram as defined by
Marquet (1994). It is close to Pearce's (1978) 
diagram except now there are two large-scale
kinetic-energy components $KS$ and $KZ$,
both connected to $KE$.
\label{FigAHCYCLE}}
\end{figure}

The six components of the cycle (\ref{eq:cycle2})
can be rearranged in various ways. An example is 
shown on Fig.~\ref{FigAHCYCLE} where the 
$A3+K2$ global cycle of Pearce is depicted 
in Fig.~\ref{FigAHCYCLE}(a) and where 
the global cycle Fig.~\ref{FigAHCYCLE}(b) is
a straightforward generalisation to a
$A3+K3$ version. 
The difference between
Fig.~\ref{FigAHCYCLE}(a) and (b) is a partitioning into
the $KZ$ and $KS$ reservoirs and the appearance of
corresponding new conversions and boundary terms. The
external path\footnote{\color{blue} 
The idea of an ``external path'' separated from 
the ``Lorenz's internal cycle'' was suggested 
in an internal report of students of the
French School of Meteorology (I. Bernard-Bouissi�res, 
M. Cadiou, A. Muzellec and Ch. Vincent, 1991), 
directed by Marc Pontaud.},
controlled by $\overline{\omega}$
(grey arrows) and corresponding to possible
large values for $AS \leftrightarrow KS$, 
is now separated from the smaller values
observed in the ``Lorenz internal cycle''
involving $AZ$, $AE$, $KZ$ and $KE$.

On one hand this version of 
Fig.~\ref{FigAHCYCLE}(b) can be 
relevant to the study of tropical cyclone 
development in regions of weak meridional 
gradients. In that case the conversions
$AS \rightarrow AZ$ and $KS \rightarrow KZ$ 
cancel out and direct transformations must
occur from $AS$ into $AE$ and  $KS$ into
$KE$. The cycle in
Fig.~\ref{FigAHCYCLE}(b) is also the one chosen
in P78 to study the energetics of dry and 
moist local convection.

On the other hand direct conversions between the 
larger-scale and eddy components may be 
considered as unrealistic. This is the case 
for midlatitude baroclinic waves where
baroclinic and barotropic conversions
$AZ \rightarrow AE$ and $KZ \rightarrow KE$
corresponding to Fig.~\ref{FigAHCYCLE}(b)
are different from classical ones 
as given by Lorenz, i.e. $CA$ and $CK$ in 
Fig.~\ref{FigCLorenz}.

\begin{figure}[hbt]
\centering
(a) \hspace*{2mm} The global $A_h$-cycle 
\hspace*{3 cm} 
(b) \hspace*{2mm} the local $a_h$-cycle
\vspace*{1mm}
\\
\includegraphics[width=0.505\linewidth,angle=0,clip=true]{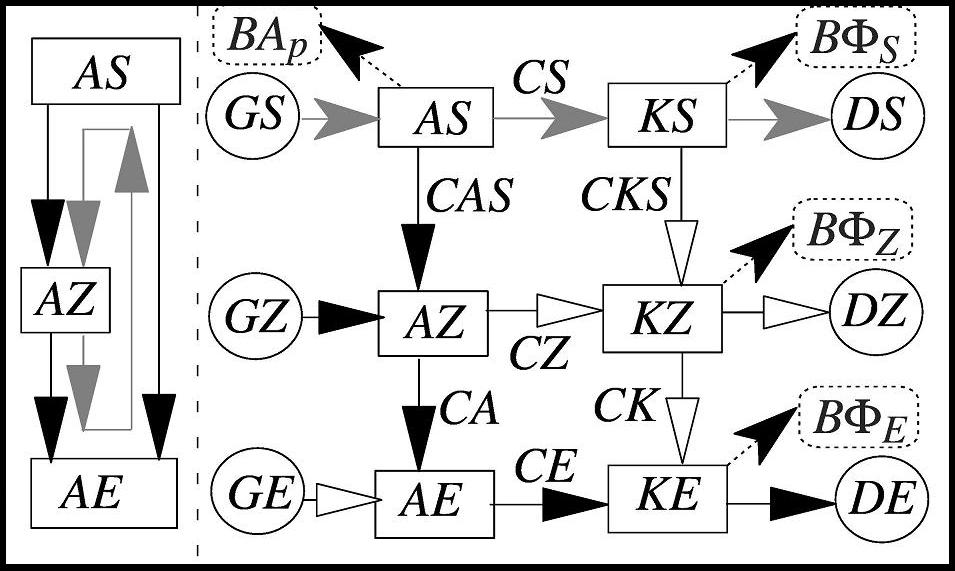}
\includegraphics[width=0.485\linewidth,angle=0,clip=true]{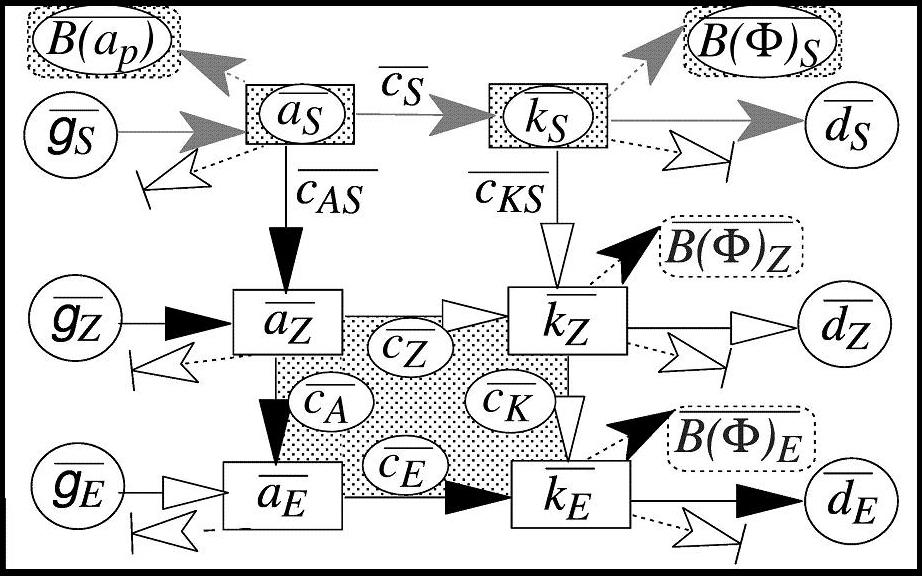}
\vspace*{-2mm}
\caption{\it \small (a) The global and symmetric available-enthalpy
cycle diagram as proposed in this paper, including the
new connections with potential-energy reservoirs 
${B\phi}_S$, ${B\phi}_Z$ and ${B\phi}_E$ (these terms do not
disappear because only the sum $\overline{B(\phi)}$ 
cancels out on global average, not individual components).
The left of the diagram shows the modification from
Fig.~\ref{FigAHCYCLE}(b): a closed 
inner loop (stippled) is formed by the
conversion term $AE \rightarrow AS$ which is
subtracted from $AS \rightarrow AE$ 
and added to the other branches 
$AS \rightarrow AZ$ and $AZ \rightarrow AE$.
A similar modification is done for the path
$KS \rightarrow KE$ and $KS \rightarrow KZ \rightarrow KE$.
The external path for energy ($AS \leftrightarrow KS$)
is now clearly separated from the internal Lorenz cycle
($AZ$, $AE$, $KZ$, $KE$) and, by comparison with
Pearce's formulation, more usual values are obtained 
for the mathematical expressions of the conversion 
terms $CA$ and $CK$.
(b) The limited area version of (a). The global 
boundary terms ${B\phi}_S$, ${B\phi}_Z$ and ${B\phi}_E$
in (a) correspond here to the ``pressure levels'' boundary 
fluxes $\overline{B(\phi)}_S (p)$, 
$\overline {B(\phi)}_Z (p)$ and $\overline {B(\phi)}_E (p)$.
The six non-labelled outgoing white arrows represent
additional boundary fluxes for each of the six energy 
components (the formulations are given in the first
terms on the right-hand sides of (\ref{eq:cycle2})).
\label{FigAHNEWCYCLE}}
\end{figure}

For these reasons, a modified version of the 
global cycle has been considered in 
Fig.~\ref{FigAHNEWCYCLE}(a) 
(the three connections with potential 
energy ($\overline{{B(\phi)}_S}$,
$\overline{{B(\phi)}_Z}$ and $\overline{{B(\phi)}_E}$)
are not shown for sake of clarity).
The modification is obtained without loss of 
generality by subtracting two internal closed loops
(depicted by grey arrows on the left part of
Fig.~\ref{FigAHNEWCYCLE}(a)),
in order to suppress the direct 
conversions $AS \rightarrow AE$ and 
$KS \rightarrow KE$. 
It has been found with 
this new version that the formulation
of the baroclinic and barotropic conversions 
$AZ \rightarrow AE$ and $KZ \rightarrow KE$
are the same as in the previous local applications 
of L55. To go from Fig.~\ref{FigAHCYCLE}(b) 
to Fig.~\ref{FigAHNEWCYCLE}(a), the conversion
$AS \rightarrow AE$ is just added to $AS \rightarrow AZ$
and $AZ \rightarrow AE$. The same is done for the 
corresponding kinetic-energy conversion terms.

The complete limited-area and pressure-level cycle 
(\ref{eq:cycle2}) corresponds to 
Fig.~\ref{FigAHNEWCYCLE}(b) where all the terms
are depicted. The boundary transport of energy
is surrounded by dashed boxes, with a shaded 
internal Lorenz cycle ($\overline{c_Z}$,
$\overline{c_A}$, $\overline{c_K}$, $\overline{c_E}$)
and with a large external path of energy:
$\overline{B(a_p)} \leftrightarrow \overline{c_S} 
\leftrightarrow \overline{{B(\phi)}_S}$.

       \subsection{Mathematical expressions for all terms.} 
      \label{subsection_4.4}

The six equations in the cycle (\ref{eq:cycle2}) are
expressed in a common form (\ref{eq:cycle2ex})
valid for an energy component $e_X$.
The time derivative $\overline{{\partial}_t(e_X )}$ is 
equal to boundary flux terms $-\overline{B( e_X )}$,
possibly with further complementary flux 
$-\overline{B(e_{cX})}$. The conversions terms
are $\pm \overline{c_A}$ and $\pm \overline{c_X}$.
The conversion of potential energy into $e_X$ is
$-\overline{{B(\phi)}_X}$, if $e_X$ is one
of the kinetic energy components. Generation 
or dissipation terms  
($+ \; \overline{g_X} \; \mbox{or} \; - \overline{d_X} $)
are present for the case of available-enthalpy
or kinetic-energy components, respectively.
\vspace{-0.15cm}
\begin{eqnarray}  
    \overline{ {\partial}_t ( e_X )  }  
      \! \!  & = &  \! \! 
               - \; \overline{  B( e_X  ) } 
            \; - \; \overline{  B( e_{cX} ) } \;\;
            \; \pm \; \overline{  c_A }       \;\;
            \; \pm \; \overline{  c_X  }      \;
            \; - \; \overline{{B( \phi )}_X}  \;\;\;
               +
    \left(  \; \overline{g_X} 
               \;\; \mbox{or}
            \; - \overline{d_X} \; \right)
            .
    \label{eq:cycle2ex}
\end{eqnarray}

The general boundary operator $\overline{B(\ldots)}$ is 
defined using (\ref{defBeta}). The special case for
$\overline{B(a_p)}= R \: T_r \: \overline{B(p)} / p$ is
obtained by using $\overline{B(p)} = \overline{\omega}$,
to give
\vspace{-0.15cm}
\begin{eqnarray}  
   \overline{ B ( a_p ) }  
   \! \! & = & \! \!
    - \: \frac{R}{p}  \; \overline{\omega} \; T_r
   \: . \label{defBap}
\end{eqnarray}

The first set of conversion terms refers to 
a transformation of any of the available-enthalpy 
reservoirs into the corresponding kinetic-energy 
component with the same status ($S$, $Z$ or $E$). 
Therefore
\vspace{-0.15cm}
\begin{eqnarray}  
   \overline{c_S}
   \! \! & = & \! \! 
   - \: \frac{R}{p}  \; \overline{\omega} \; \overline{T}
   \: , \hspace{0.6cm}
   \overline{c_Z}
   \; = \;
   - \: \frac{R}{p}  \; \overline{ {\omega}^{\lambda}_{\varphi}  
                       \; {T}^{\lambda}_{\varphi} } 
  \: , \hspace{0.6cm}
   \overline{c_E}
   \; = \;
   - \frac{R}{p}  \;  \overline{ {\omega}_{\lambda} 
                       \; {T}_{\lambda} }  
  \: , \label{defCSCZCE} 
\end{eqnarray}
where the baroclinic conversions $\overline{ c_Z}$ and $\overline{ c_E}$
take the classic form. Note that there is no 
implicit summation over repeated ${\lambda}$ or 
${\varphi}$ subscripts or superscripts.

The second set of conversion terms represents energy
transformations from one form to another between the
three available-enthalpy reservoirs or the
three kinetic-energy reservoirs. Thus
\begin{eqnarray}
   \overline{c_{AS}}  
    & = & 
   - \: c_p \; \overline{( {\omega}^{\:\prime}  \; 
                                {T}^{\:\prime} )} 
   \; \; p^{- {\kappa}} \; \frac{\partial}{\partial \, p} 
   \left\{ p^{\kappa} \left(1 - \frac{T_r }{\overline{T}} \right) \right\}
   \: , \label{defCAS}
 \\
   \overline{c_{KS}}  
    & = & 
   -     
   \left\{ \: 
   \overline{( {u}^{\:\prime}  \; {\omega}^{\:\prime} )} \; 
   \;  \frac{\partial \,\overline{u}}{\partial \, p} 
   \; + \; 
   \overline{( {v}^{\:\prime}  \; {\omega}^{\:\prime} )} \; 
   \;   \frac{\partial \,\overline{v}}{\partial \, p} 
   \: \right\}
   \: , \label{defCKS} 
 \\
   \overline{c_A}  
    & = & 
   - \: c_p
     \left[
     \; \overline{
     {({v}_{\lambda} \: {T}_{\lambda} )}^{\lambda}
     \; \frac{\partial}{\partial \, y} 
    \left(   1 -  \frac{T_r }{{T}^{\lambda}}
    \right) }
   \; +
     \; \overline{
     {({\omega}_{\lambda} \: {T}_{\lambda} )}^{\lambda}
     \: p^{- {\kappa}} \: 
     \frac{\partial}{\partial \, p} 
     \left\{    p^{\kappa} \:
      \left(   1 - \frac{T_r }{ {T}^{\lambda}} \right)
     \right\} }
   \: \right]
   , \label{defCACK} 
 \\
  \overline{c_K}    
    & = & 
   -     
   \left\{ \: 
   \overline{ 
       {({u}_{\lambda}  {v}_{\lambda})}^{\lambda}   \: 
      \frac{\partial \,{u}^{\lambda}}{\partial \, y} 
   }
   + 
   \overline{ 
       {({v}_{\lambda}  {v}_{\lambda})}^{\lambda}   \: 
       \frac{\partial \,{v}^{\lambda}}{\partial \, y} 
   }
   + 
   \overline{ 
       {({u}_{\lambda} \: {\omega}_{\lambda})}^{\lambda}   \: 
       \frac{\partial \,{u}^{\lambda}}{\partial \, p} 
   }
   + 
   \overline{ 
       {({v}_{\lambda} \: {\omega}_{\lambda})}^{\lambda}   \: 
      \frac{\partial \,{v}^{\lambda}}{\partial \, p} 
   }    
   \: \right\}
   . \label{defCK} 
\end{eqnarray}
The baroclinic and barotropic conversions 
$\overline{c_A}$ and $\overline{c_K}$
take the classical form.

The boundary terms 
$\overline{{[B(\phi)]}{}_{\textstyle \eta}}$
with $\eta = S, Z$ or $E$ are the
projections of $\overline{B(\phi)}$
onto the three equations
for $k_S$, $k_Z$ and $k_E$, respectively.
They can be interpreted as conversion terms with 
the potential energy because $\overline{B(\phi)}$
appears with opposite signs in equations for
kinetic and potential energies (\ref{defevolek}) 
and (\ref{defevolPhi}).
They cannot be put in a form 
$\overline{B[{\phi}_{\textstyle \eta}]}$,
i.e. the boundary flux of some
${\phi}_{\textstyle \eta}$ to be determined,
as indicated in the local studies of 
Muench (1965), Brennan and Vincent (1980)
or Michaelides (1987).
As a consequence
$\overline{{[B(\phi)]}{}_{ \textstyle \eta}} \neq 0$,
with $\eta = S, Z$ or $E$.
These terms could not appear in the global
approaches of L55 or P78 because the sum of the 
three projections, i.e. $\overline{B(\phi)}$, 
has been cancelled out at the beginning 
of these studies, as a global term equal to zero.
However, even in L55 and P78, the terms 
$\overline{{[B(\phi)]}{}_{ \textstyle \eta}}$
is different from $0$ and they should have been present.

The other boundary term $\overline{B(a_p)}$
with $a_p = R \: T_r \; \ln(p/p_r)
  = R \: T_r \; \{ \ln(p) - \ln(p_r) \}$
cancels out if it is integrated over $p_t$ and
$p_b$. It is a consequence of the definition 
of $p_r$ when $\ln(p_r)$ is equal to the global 
average of $\ln(p)$.
\vspace*{-0.15cm}
\begin{align} 
  \overline{{B( \phi )}_S} 
   \; & = \; 
   \overline{ u }    \; \overline{ \left(\frac{\partial \, \phi}{\partial \, x}\right)  } \: + \:  
   \overline{ v }    \; \overline{ \left(\frac{\partial \, \phi}{\partial \, y}\right)  } \: + \:  
   \overline{\omega} \; \overline{ \left(\frac{\partial \, \phi}{\partial \, p}\right)  } 
   \hspace{11mm} \: = \: 
     \overline{{\bf U}_h} \: . \: 
     \overline{{\bf \nabla}_{\!p} \, \phi }
  \:  - \: \frac{R}{p} \; \overline{\omega} \: \overline{T}
   \: , \label{defBPHIS} 
 \\
    \overline{{B( \phi )}_Z}
   \; & = \; 
   \overline{ { u }^{\lambda}_{\varphi}    
             \left( \frac{\partial \, \phi}{\partial \, x}\right)
     \hspace*{-5.4mm}{\phantom{\Bigl( \Bigr)}}^{\lambda}_{\varphi} }  \: + \: 
   \overline{ { v }^{\lambda}_{\varphi}    
             \left( \frac{\partial \, \phi}{\partial \, y} \right)
     \hspace*{-5.4mm}{\phantom{\Bigl( \Bigr)}}^{\lambda}_{\varphi} }  \: + \:  
   \overline{ {\omega}^{\lambda}_{\varphi} 
             \left( \frac{\partial \, \phi}{\partial \, p} \right)
     \hspace*{-5.4mm}{\phantom{\Bigl( \Bigr)}}^{\lambda}_{\varphi} }  
   \hspace{0.5mm} \: = \:
     \overline{{({\bf U}_h)}^{\lambda}_{\varphi} \: . \: 
     {({\bf \nabla}_{\!p} \,  \phi)}^{\lambda}_{\varphi}}
   \: - \: \frac{R}{p}  \; \overline{ {\omega}^{\lambda}_{\varphi}  
     \: {T}^{\lambda}_{\varphi} }
   \: , \label{defBPHIZ}
 \\
  \overline{{B( \phi )}_E}  
   \; & = \; 
   \overline{ { u }_{\lambda}    
             \left( \frac{\partial \, \phi}{\partial \, x} \right)
     \hspace*{-5.4mm}{\phantom{\Bigl( \Bigr)}}_{\lambda} } \: + \:   
   \overline{ { v }_{\lambda}    
             \left(  \frac{\partial \, \phi}{\partial \, y} \right)
     \hspace*{-5.4mm}{\phantom{\Bigl( \Bigr)}}_{\lambda} } \: + \:  
   \overline{ {\omega}_{\lambda} 
             \left(  \frac{\partial \, \phi}{\partial \, p} \right)
     \hspace*{-5.4mm}{\phantom{\Bigl( \Bigr)}}_{\lambda} }
   \hspace{2mm} \: = \:
     \overline{{({\bf U}_h)}_{\lambda} \: . \: 
     {({\bf \nabla}_{\!p} \,  \phi)}_{\lambda}}
   \: - \: \frac{R}{p}  \; \overline{ {\omega}_{\lambda} \: {T}_{\lambda} } 
   \: . \label{defBPHIE}
\end{align}

Note that the baroclinic conversions $\overline{c_S}$,
$\overline{c_Z}$ and $\overline{c_E}$ appear with the 
opposite sign in the conversion terms 
$\overline{c(\phi,k_X)} = - \overline{{B(\phi)}_X} $
for $X=(S,Z,E)$ in Equations (\ref{defBPHIS}) to (\ref{defBPHIE}).
This unexpected property will be discussed in more detail
in part~II of this paper. The result is that these
combinations of terms are equal to the work of the general 
pressure forces 
`$- {\bf \nabla}_{\!p} (\phi)$'
against the motion and after a projection onto
the subset $X=(S,Z,E)$ of components.
It gives rise to the equations
\vspace*{-0.15cm}
\begin{eqnarray} 
     \overline{c_S} \; - \; \overline{{B(\phi)}_S}
     & = & - \;
     \overline{{\bf U}_h} \: . \: 
     \overline{{\bf \nabla}_{\!p} \phi }
   \: , \label{defCPLUSBPHIS} 
 \\
     \overline{c_Z} \; - \; \overline{{B(\phi)}_Z}
     & = & - \;
     \overline{{({\bf U}_h)}^{\lambda}_{\varphi} \: . \: 
               {({\bf \nabla}_{\!p} \phi)}^{\lambda}_{\varphi}}
   \: , \label{defCPLUSBPHIZ}
 \\
     \overline{c_E} \; - \; \overline{{B(\phi)}_E}
     & = &  - \;
     \overline{{({\bf U}_h)}_{\lambda} \: . \: 
               {({\bf \nabla}_{\!p} \phi)}_{\lambda}}
   \: . \label{defCPLUSBPHIE}
\end{eqnarray} 

Finally, the generation and dissipation terms are written
as follows
\vspace*{-0.15cm}
\begin{eqnarray}
   \overline{ g_S }  
   \! \! & = & \! \!
   \left( \: 1 -  \frac{T_r}{\overline{T}} \: \right) \; \overline{(\:\dot{q}\:)}
   \hspace{2.5cm} ; \hspace{0.4cm}
   \overline{ d_S }
   \; \; \;  = \; \; \;
   - \: \overline{{\bf U}_h} \: . \: \overline{{\bf F}_h}
   \label{eq:ets67}
   \: , \label{eq:ets93}
 \\
   \overline{ g_Z }  
   \! \! & = & \! \!
    \left(\: \frac{T_r}{\overline{T}} \:\right) \;
   \overline{ \left\{ \: \left( \: 1 -  \frac{\overline{T}}{{T}^{\lambda}} \: \right) 
                \; \dot{q}  \: \right\} }
   \hspace{0.7cm} ; \hspace{0.4cm}
   \overline{ d_Z }  
   \; \; \;  = \; \; \;
   - \: \overline{ {({\bf U}_h)}^{\lambda}_{\varphi} } \: . \: 
        \overline{ {({\bf F}_h)}^{\lambda}_{\varphi} }
   \: , \label{eq:ets94}
 \\
   \overline{ g_E }  
   \! \! & = & \! \! 
   \overline{ \left\{ \: \left(\: \frac{T_r}{{T}^{\lambda}} \:\right)  \; 
              \left( \: 1 -  \frac{{T}^{\lambda}}{T} \: \right) 
                \; \dot{q}  \: \right\} }
   \hspace{0.6cm} ; \hspace{0.4cm}
   \overline{ d_E }
   \; \; \;  = \; \; \;
   - \: \overline{ {({\bf U}_h)}_{\lambda} } \: . \: 
        \overline{ {({\bf F}_h)}_{\lambda} }
   \: . \label{eq:ets95}
\end{eqnarray}
It is possible to develop the Carnot 
factors in the generation terms $\overline{ g_Z }$
and $\overline{ g_E }$ in order to bring the
expressions closer to the corresponding values given
in P78: 
$\overline{ \{
{T}^{\lambda}_{\varphi} \: {(\dot{q})}^{\lambda}_{\varphi} 
\} } / T_r$ 
and $\overline{ \{
{T}_{\lambda} \: {(\dot{q})}_{\lambda}
\} } / T_r$, 
respectively. 
The detailed computations are presented in Appendix-C
and the final approximate formulae are given by
(\ref{AppC:etsGZapprox}) and 
(\ref{AppC:etsGEapprox}). 
The results
\begin{eqnarray}
  \overline{g_Z} & \approx &
 \overline{
   \: \left\{ 
     \frac{
      T^{\lambda}_{\varphi}
      \; {(\dot{q})}^{\lambda}_{\varphi}
      }
     {{T}^{\lambda}} 
     \right\}
 } 
 \: \left( \frac{T_r}{\overline{T}} \right)
\nonumber \\
 \overline{g_E} & \approx &
 \overline{ 
   \left\{ \: 
     { \left\{ 
     \frac{T_{\lambda} \: {(\dot{q})}_{\lambda}}
      {T}
      \right\} 
      }^{\lambda}
     \: \left( \frac{T_r}{{T}^{\lambda}} \right) 
  \right\}
 }
\nonumber
\end{eqnarray}
compare relatively well with P78's expressions.

\section{\Large \underline{A modified temporal scheme}.}
 \label{section_5}

An output dataset from the French Arpege model
will be used in part~II of this paper
for post-processed data on $27$ pressure levels at uneven 
intervals from $10$ to $1000$~hPa. 
The time derivative 
and other terms of the cycle (\ref{eq:cycle2}) will 
be evaluated with meteorological data, available every $3$~hours. 
The questions to be addressed are: 
(i) How to compute the time and spatial differencing? 
(ii) What are the accuracies of these schemes? 

It is usual in papers dealing with energetics 
to express the time derivative at time $t_0$ 
as a centred finite-difference scheme
computed between $ t_{(-)} = t_0 - \Delta t$ 
and $ t_{(+)} = t_0 + \Delta t$.
If (\ref{eq:cycle2}) is 
schematically represented by 
${\partial}_t (Z) = C$
and if the notations $ Z^{(+)}$, $ Z^{(-)}$ 
and $ C^{(0)}$ are used for values at time 
$ t_{(+)} $, $ t_{(-)} $ and $t_0$, 
the usual scheme can be rewritten as
\vspace{-0.15cm}
\begin{eqnarray}
\frac{ Z^{(+)} - Z^{(-)}}{2 \;\Delta t} &  
\approx & C^{(0)} \; \; \equiv \; \;
\{ {\partial}_t (Z) \}{}^{(0)} \: .
\label{eq:bilanDER}
\end{eqnarray}

An objective evaluation of the quality of this 
scheme will be done using the test functions 
$Z(t) = \cos(\omega t)$ and 
$C(t)=-\omega\:\sin(\omega t)$, where
$\omega = 2 \pi / T_0$. If $C^{(0)}$ is taken
as a reference value,
the approximation (\ref{eq:bilanDER}) corresponds to 
$\{\sin(\pi \ell)\}/(\pi \ell) \approx 1$
for $\ell = \Delta t / {(\Delta t)}_{crit}$
and for a critical time interval equal to
${(\Delta t)}_{crit} =  T_0/2 = \pi/\omega$. 
In that case, the relative error is equal to
${\varepsilon}_1 = 1-\{\sin(\pi \ell)\}/(\pi \ell)$.
Values of ${\varepsilon}_1$ are given 
for $\ell = 0.1$ to $\ell = 1.4$ in
Table~{\ref{TabNEWSCHEME}}. This scheme becomes rapidly
inaccurate for $\ell > 0.3$ or equivalently
for $\Delta t > {(\Delta t)}_{crit}/3$, 
with errors reaching $36$~\% or more when 
$\ell>0.50$.

\begin{table}[htb]
\centering
\caption{\it \small The error functions.
Values of ${\varepsilon}_1 = 1-\{\sin(\pi \ell)\}/(\pi \ell)$ 
and ${\varepsilon}_2 = {\varepsilon}_1 + \{\cos(\pi \ell)-1\}/3$
for $\ell = \Delta t / {(\Delta t)}_{crit}$ and $\; \: 0.1 \leq \ell \leq  1.4$
\label{TabNEWSCHEME}}
\vspace*{2mm}
\centerline
{\begin{tabular}{|c|ccccccccccccc|}
\hline 
\hline 
$\ell$             & 0.1  & 0.2  & 0.3  & 0.4  & 0.5  & 0.6  & 0.7  & 0.8  & 0.9 & 1.0 & 1.2 & 1.4 &\\ 
\hline 
\hline 
 ${\varepsilon}_1$ & 0.02 & 0.06 & 0.14 & 0.24 & 0.36 & 0.50 & 0.63 & 0.77 &   0.89 &  1.0 & 1.16 & 1.22 &\\ 
\hline 
 ${\varepsilon}_2$ & 0.00 & 0.00 & 0.00 & 0.01 & 0.03 & 0.06 & 0.10 & 0.16 &   0.24 &  0.33 & 0.55 & 0.78 &\\ 
\hline 
\hline 
\end{tabular}}
\end{table}

The proposal of this present paper is to write an
improved scheme that could manage cases when
$\Delta t \approx {(\Delta t)}_{crit}$.
The new approach is based on an approximate
form of the integral of $C(t)$ in the interval
$[ t_{(-)} ;  t_{(+)} ]$ and on
an exact result for the integral of
${\partial}_t (Z)$.
\vspace{-0.15cm}
\begin{eqnarray}
   \int_{t_{(-)}}^{t_{(+)}}
       \!\!  {\partial}_t (Z) \: dt
\; \equiv \;
   Z^{(+)} - Z^{(-)}
\! & = & \!
   \int_{t_{(-)}}^{t_{(+)}}
        \!\!  C \: dt 
\; \; \; \Longrightarrow \nonumber
\\ \frac{1}{2 \; \Delta t} \;
   \int_{t_{(-)}}^{t_{(+)}}
        \!\!  C \: dt
\; = \;
   \frac{ Z^{(+)} - Z^{(-)} }
   {2 \; \Delta t}
\! & \approx & \!
   C^{(0)} \; + \;  
   \frac{  \: C^{(+)} -2 \, C^{(0)} + C^{(-)} \: }{6}
\; .  \label{eq:bilanINT}
\end{eqnarray}
The integral is computed with an approximation 
of $C(t)$ around $t=t_0$ by a quadratic function
of time, defined as $C(t) = a (t-t_0){}^2 + b(t-t_0) +c$.
The three constants are determined by
$C(t_{(-)}) = C^{(-)}$, $C(t_0) = C^{(0)}$
and $C(t_{(+)}) = C^{(+)}$. They are equal to 
$a=\{  \:  C^{(+)} -2 \, C^{(0)} + C^{(-)} \:
\}/ \{ 2 {(\Delta t)}^2 \} $,
$b= \{C^{(+)} - C^{(-)} \}/(2 \Delta t) $ 
and $c=C^{(0)}$. 
As a result the old scheme (\ref{eq:bilanDER}) 
is transformed into (\ref{eq:bilanINT}).
There is an additional term
with weighting factors ($1/6$, $-1/3$, $1/6$).
It is zero when $C(t)$ varies linearly with time 
but can be large in the case of rapid increase or decrease 
of its change in time.

The objective evaluation for (\ref{eq:bilanDER}) 
can also be realized for (\ref{eq:bilanINT}) 
although it leads to the relative error 
${\varepsilon}_2 = {\varepsilon}_1 + \{ \cos(\pi \: \ell) -1 \}/3$.
The corresponding limits of ${\varepsilon}_1$ and 
${\varepsilon}_2$ for small $\ell$ are ${(\pi \: \ell)}^2/6$ and 
${(\pi \: \ell)}^4/180$, respectively. The accuracy is clearly 
improved and the second-order scheme becomes a 
fourth-order scheme. Numerical values of ${\varepsilon}_2$ 
are given in Table~{\ref{TabNEWSCHEME}} and it appears
that the new scheme is accurate enough up to ${\ell}<0.8$,
with an error for $\Delta t \approx {(\Delta t)}_{crit}$ 
decreasing from $100$~\% to $33$~\% when replacing 
(\ref{eq:bilanDER}) by (\ref{eq:bilanINT}).
\begin{figure}[hbt]
\centering
(a) \hspace*{2mm} Slow advection ($0.7\:L$)
\hspace*{4 cm} 
(b) \hspace*{2mm} Rapid advection ($1.2\:L$)
\\
\includegraphics[width=0.48\linewidth,angle=0,clip=true]{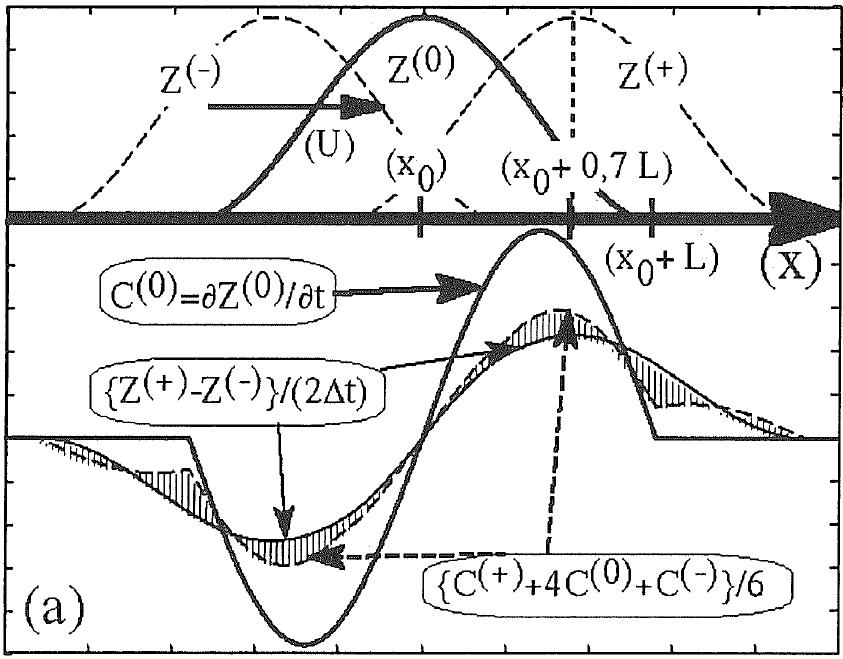}
\includegraphics[width=0.48\linewidth,angle=0,clip=true]{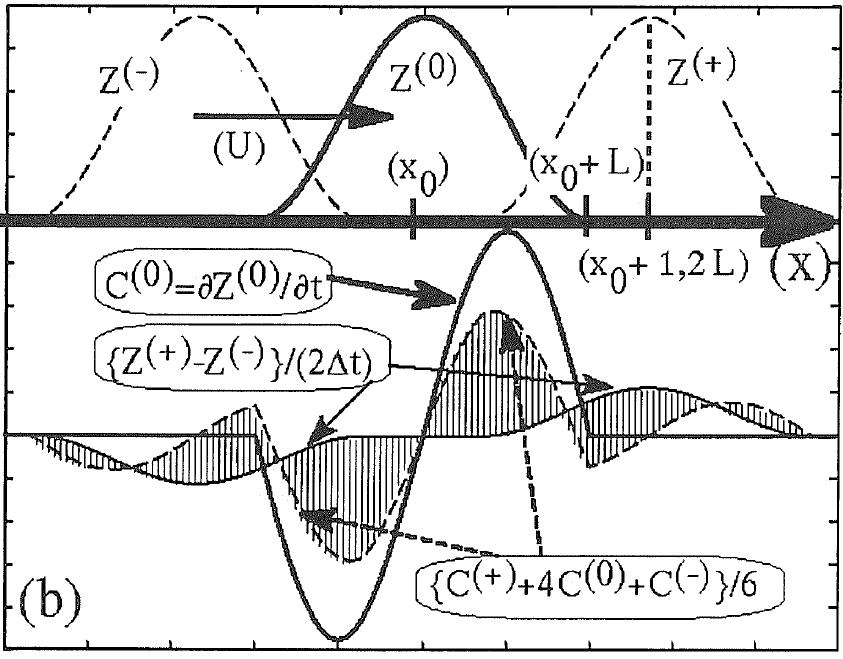}
\vspace{-2mm}
\caption{\it \small 
A comparison of accuracy for different 
numerical schemes to solve ${\partial}_t (Z) = C$ for 
(a) moderate advection and for
(b) more rapid advection.
The pattern $Z=\cos\{\pi(x - x_0)/L\}$ where 
$|x - x_0|<L$, is advected by wind $U$ with 
$x_0 = U (t-t_0)$. Length $L$ corresponds the
pattern radius. On the upper parts of 
panels (a) and (b) the curves $Z(x)$ are shown
at times $t_0 - \Delta t$, $t_0$ and $t_0 + \Delta t$.
On lower parts the true time derivative $C^{(0)}$ is 
depicted by a heavy solid line. A thin solid line shows the
approximate finite difference (\ref{eq:bilanDER}) and
a dashed line shows the new scheme (\ref{eq:bilanINT}).
For (a) and (b), the true time derivative $C^{(0)}$
cannot be compared with $\{ Z^{(+)} - Z^{(-)}) \} 
/ (2 \Delta t)$ and (\ref{eq:bilanDER}) 
is not verified. 
It is also clear that (\ref{eq:bilanINT})
is valid for (a) because the dashed and
thin solid lines are close together, 
encompassing a small hatched area.
But for a rapid advection case (b) the new scheme 
is no longer valid, as indicated by a hatching area
which is too large.
It should be mentioned that for small advection 
(say $\ell=0.3$, not shown) 
the dashed and thin solid lines cannot be distinguished 
and (\ref{eq:bilanINT}) is almost exact. Furthermore,
for $\ell=0.3$ the solid line is already separated 
from the others, which means that (\ref{eq:bilanDER}) 
is rapidly not verified for increasing $\ell$, even 
when $\Delta t \ll {(\Delta t)}_{crit}$.
Symbols are explained in Appendix A.
\label{FigNEWSCHEME}}
\end{figure}

Improvements in scheme accuracy are
confirmed on the basis
of the subjective visual analyses 
presented in Fig.~(\ref{FigNEWSCHEME}).
Let us consider the advection by a uniform
wind $U$ of a pattern defined by
$Z(x,t)=\cos(k x - \omega t + {\alpha}_0)$,
with $k=\pi / L$, $\omega = \pi U / L$ and 
${\alpha}_0 = \pi U t_0 / L$.
The accuracies of the old and new schemes 
(\ref{eq:bilanDER}) and (\ref{eq:bilanINT}) can be 
evaluated by visual comparisons of the different lines 
depicted on the lower parts of 
Fig.(\ref{FigNEWSCHEME})(a) and
(b), with $\ell \approx 0.7$ for 
Fig.(\ref{FigNEWSCHEME})(a) and 
$\ell \approx 1.2$ for 
Fig.(\ref{FigNEWSCHEME})(b). 

It appears that the old scheme is not verified for 
any of the two cases because the heavy lines for $C^{(0)}$
are clearly separated from the other two. 
The accuracy of the new scheme can
be appreciated by the small hatched area.
If the new scheme is valid for the moderate advection 
scheme in Fig.(\ref{FigNEWSCHEME})(a), 
it is no longer valid in Fig.(\ref{FigNEWSCHEME})(b) 
when $\Delta t \approx 1.2 {(\Delta t)}_{crit}$.
So, the subjective limits are equivalent to the objective ones
and  in order to ensure an accuracy better than $16$~\% 
the time interval must verify $\ell < 0.3$ for
(\ref{eq:bilanDER}) and $\ell < 0.8$ for (\ref{eq:bilanINT}).

The critical time interval 
can be small in the case of real small-scale 
meteorological features like frontal 
waves or mobile troughs. 
For the example, presented in
Figure (3) of Michaelides (1987), the observation of
the successive panels indicate a radius of about $10\,{}^{\circ}$
for the depression and a zonal advection of about 
$10\,{}^{\circ}$~{(day)}${}^{-1}$. As a consequence,
the critical time interval is equal to $1$~{day} and
the limits required for relevant applications of 
(\ref{eq:bilanDER}) or (\ref{eq:bilanINT}) are 
respectively equal to $7$~{h} and $19$~{h}.
 
The new scheme can be interpreted 
as a `moving average' approach centred on $t_0$,
with a window of $\pm \Delta t$.
It is when all terms in (\ref{eq:cycle2}) are computed
with (\ref{eq:bilanINT}) as moving averages
that the dissipation and generation terms can be
derived as moving average residuals. Large
values of dissipation and generation terms, described 
in previous papers dealing with local energetics,
are perhaps partly due to the imbalance in (\ref{eq:bilanDER})
and to values of $\Delta t$ close to or above
${(\Delta t)}_{crit}$.

\section{\Large \underline{Discussion of $T_r$, $p_r$ and the reference state}.} 
 \label{section_6}

The choice of prescribed and constant values for $T_r$ and $p_r$
is often open to criticism. It appears to be a problem for the
use of more complex reference states (non-uniform, 
non-stationary or without zonal-mean symmetry).
However, the possibility of  choosing more complex 
``reference states'' does exist with the present
available-enthalpy approach. There is a real possibility of 
defining a less academic  and more realistic basic state.

Even if a constant temperature  $T_r$ is used as in P78
to define an isothermal ``thermodynamic reference atmosphere'',
in Pearce's and this paper,
the real ``reference meteorological state'' must be 
thought of in terms of the additional isobaric and zonally averaged 
quantities $\overline{T}(p, t)$ and ${T}^{\lambda}(\varphi, p, t)$.
These are time-dependent states of the atmosphere, based on real
meteorological datasets.

Another isothermal reference state $T_0=\:$constant have been used 
by Andrews (1981) to define the potential energy 
for a perfect gas by
\begin{equation}
\Pi \; = \; {\Pi}_1(p/p_0)
\: + \:
{\Pi}_2(\theta/{\theta}_0)
\nonumber \: ,
\end{equation}
where
the two parts ${\Pi}_1$ and ${\Pi}_2$ are 
local and positive-definite everywhere. 
The second part can be written as
\begin{equation}
{\Pi}_2(\eta) \; = \; c_p \; T_0 \; h(\eta)
\nonumber \: ,
\end{equation}
with $h(\eta) = \exp(\eta) -1 - \eta$
and $\eta=\ln(\theta/{\theta}_0)$.
It is thus equal to
\begin{equation}{\Pi}_2(X) \; = \; c_p \; T_0 \; {\cal F}(X)
\nonumber \: ,
\end{equation}
where $X=\theta/{\theta}_0-1$ and the function ${\cal F}(X)
= X - \ln(1+X)$ is the same function
used  in (\ref{defFX}) to define $a_T = c_p \: T_r \; {\cal F}(T/T_r-1)$.

The reference values $T_r$ and $p_r$ are introduced in order 
to isolate the pressure component 
$a_p  = R  \;  T_r \: \ln ( p / p_r )$ 
and to define the ``zero-order'' quadratic function
$a_T  = c_p \: T_r \:{\cal F}\{ ( T - T_r ) / T_r \}$.
The separating property (\ref{propFX1X2}) has then been
used to successively insert the first- and second-order 
departure terms of
Lorenz: $T_1={T}^{\lambda}_{\varphi}$ 
and $T_2={T}_{\lambda}$.
The same separating property can easily be used to deal with
other ``eddy'' and ``mean'' energetic investigations, with
more complex definitions for the average terms (e.g.
for temporal or spatial moving averages of the flow,
with possible tilted features and complex geometry over
the limited area domain).

Actually the choice of $T_r$ and $p_r$ is
not a central point for atmospheric purposes. Other
developments could be pursued in the future with
other possible definitions for $T_1$ and $T_2$ 
to be inserted between the same $T$ and $T_r$.
A new problem would, however, appear in that case since
the algebra in Appendix B would not be easy to
use, making the derivation of the energy cycle 
(\ref{eq:cycle2}) very difficult in most cases.
Here lies the success of Lorenz's separation
into ${T}^{\lambda}_{\varphi}$ 
and  ${T}_{\lambda}$, even when it is
applied to the available-enthalpy function
and to limited-area domains.

\section{\Large \underline{Conclusions}.} 
 \label{section_7}

The specific available enthalpy function (\ref{defs})
has been used to derive the complete energy cycle (\ref{eq:cycle2}).
It can be applied to any pressure level of any limited-area
atmospheric domain. It is an exact cycle with $A3+K3$ components 
corresponding to Fig.~\ref{FigAHNEWCYCLE}~(b). There are no 
approximations and no missing terms. The demonstration of this affirmation
will be obtained in part~II of this paper when the generation and 
the dissipation terms will be computed as residuals of 
Eqs.(\ref{eq:cycle2}) using the modified temporal 
scheme (\ref{eq:bilanINT}).  
Logically these residuals should be small in the case of
adiabatic studies of idealized simulations of baroclinic waves,
whereas they were very large in previous studies where some
terms of (\ref{eq:cycle2}) were missing.

The approach followed in this paper is similar to some extent
to  L55 and P78 in that the basic states
${T}^{\lambda}$, ${u}^{\lambda}$ and ${v}^{\lambda}$ are zonally 
symmetric. The eddy and zonally symmetric components
$\overline{a_E}$ and $\overline{a_Z}$ are almost the same
as in L55, but the averaged stability $\overline{\sigma}$ is 
disregarded and replaced, as in P78, 
by an additional static stability component
$a_S$. The kinetic components are
defined similarly by $\overline{k_E}$, $\overline{k_Z}$
and $k_S$, for the sake of symmetry with the partitioning
of the available enthalpy components.

The global and local (``pressure level'') versions of the 
available-enthalpy cycle represented by Fig.~\ref{FigAHNEWCYCLE}~(a) and 
(b) differ only by some boundary fluxes, as it should be.
The transformations required to go from Lorenz cycle to the
local available-enthalpy cycle are illustrated by considering 
the series of  Figs.\ref{FigCLorenz}, \ref{FigAHCYCLE}~(b), 
\ref{FigAHNEWCYCLE}~(a) and \ref{FigAHNEWCYCLE}~(b).

In the new cycle, the baroclinic and barotropic 
conversion terms $\overline{c_E}$
and  $\overline{c_Z}$ in (\ref{defCSCZCE}) and (\ref{defCK}) take 
their classical form. However, they are obtained by the usual
transformation of horizontal wind 
$- {\bf U}_h \: . \: {\bf \nabla}_{\!p} \:(\phi)$
into $- \: B(\phi) - R \: \omega \: T / {p}$, where the
second term uses the vertical wind. This manipulation
leads to cancellation of the boundary term $- \: B(\phi)$
on a global scale which simplifies the study of L55.

This, though, is no longer true for the local study 
presented here. It could be more advantageous to keep the 
initial formulation $- {\bf U}_h \: . \: {\bf \nabla}_{\!p} \:(\phi)$
when budgets of kinetic-energy components are considered,
or equivalently $ f \: {\bf k} \: . 
\left( {\bf U}_g \times {\bf U}_a \right)$, where
${\bf U}_g$ and ${\bf U}_a$ are the geostrophic and 
ageostrophic winds, respectively.
The main problem is that the classical baroclinic conversion 
$- R \: \overline{ {\omega}_{\lambda} \: {T}_{\lambda} } / p$
is contained in both $\overline{c_E}$ and in $- \overline{{B(\phi)}_E}$ 
with the opposite sign. As a consequence, it does not contribute 
to any change in $\overline{k_E}$ and a careful comparison 
of the two formulations, using ageostrophic
or vertical winds, is thus necessary.
This will be done in part~II, based on 
applications to idealize adiabatic and diabatic simulations.

\vspace{5mm}
\noindent{\Large\bf \underline{Acknowledgements}.}
\vspace{2mm}

The author is most grateful to S. Malardel for his
initial support regarding the applications of 
available-enthalpy energetics to idealized simulations. 
I also thank R. Clark and the two referees who suggested 
many clarifications and modifications to the manuscript.

\vspace{6mm}
\noindent
{\Large\bf Appendix A.    Basic notation.}
             \label{appendix_A}
\renewcommand{\theequation}{A.\arabic{equation}}
  \renewcommand{\thefigure}{A.\arabic{figure}}
   \renewcommand{\thetable}{A.\arabic{table}}
      \setcounter{equation}{0}
        \setcounter{figure}{0}
         \setcounter{table}{0}
\vspace{1mm}
\hrule

\vspace*{-1mm}
\begin{tabbing}
 ---------------------------------\=  ---------------------------------------------------------- --\= \kill
$a$, $A$  \>  Local and global available enthalpy (Pearce, 1978). \\
$a_h$, $A_h$ \> Local specific and global  available enthalpy. \\
$a_T$, $A_T$ \> Local specific and global  temperature-component of $a_h$ and $A_h$. \\
$a_p$, $A_p$ \> Local specific and global  pressure-component       of $a_h$ and $A_h$. \\
$\overline{a_S}$, $\overline{a_Z}$, $\overline{a_E}$ \>  Basic available-enthalpy components. \\
$\overline{a_{cS}}$, $\overline{a_{cZ}}$ \>  Complementary available-enthalpy components. \\
$a_{e1}$, $a_{e2}$  \>  Local specific values for two  available-energies. \\
APE \> Global available potential energies  (Lorenz, 1955) \\
$\overline{B(\ldots)}$ \> Boundary flux terms for all energy components \\
$\overline{{B(\phi)}_S}$, $\overline{{B(\phi)}_Z}$, $\overline{{B(\phi)}_E}$ \> Potential-energy special conversion terms \\
$B$, $E$, $S$, $Z$ \> Subscripts for baroclinicity, eddy, static-stability and zonal components \\
$BA_h$, $BK$, $BA_p$, $B(\phi)$  \>   Global boundary flux terms \\
$\overline{c_A}$, $\overline{c_Z}$, $\overline{c_E}$, $\overline{c_K}$ \> Basic conversions \\
$\overline{c_S}$, $\overline{c_{AS}}$,  $\overline{c_{KS}}$ \> Other basic conversions involving $\overline{a_S}$ \\
$\overline{c_a}$ \> Ageostrophic conversion (Part~II) \\
$\overline{(c_{ag})_S}$, $\overline{(c_{ag})_Z}$, $\overline{(c_{ag})_E}$ \> Ageostrophic conversions (Part~II) \\
$c_p$\> Specific heat at constant pressure for dry air\\
$d$, $D$  \>   Local and global dissipation terms \\
$\overline{d_S}$, $\overline{d_Z}$, $\overline{d_E}$  \> Dissipation terms \\
$e_i$, ${(e_i)}_r$ \>  Local specific values for  internal  energy, reference value of $e_i$  \\
$e_k$ \>  Local specific values for kinetic  energy \\
$e_p = \phi$ \>  Local specific values for potential  energy \\
$E_i$  \>  Global internal energy \\ 
$E_p$  \>  Global potential energy \\ 
$f$, $f^{\ast}$ \>   Coriolis and pseudo Coriolis factor \\
${\bf F}_h$, $(F_h)_x$, $(F_h)_y$ \> Frictional force and its horizontal components \\
$\cal{F}$ \> An exergy (quadratic) function \\
$\cal{G}$ \> An exergy (quadratic) function \\
$g$, $G$  \>   Local and global  generation terms \\
$\overline{g_S}$, $\overline{g_Z}$, $\overline{g_E}$  \> Generation terms \\
$h$, $h_r$, $h_{00}$ \> Local specific enthalpy, reference and standard values of $h$ \\
$H$ \>  Global enthalpy \\ 
$H_{pbl}$  \>   Scale height of the Planetary Boundary Layer (Part~II) \\
${\bf k}$ \> Vertical unit vector \\
$\overline{k_S}$, $\overline{k_Z}$, $\overline{k_E}$ \>  Pressure-level average kinetic-energy components of $e_k$ \\
$\overline{K_S}$, $\overline{K_Z}$, $\overline{K_E}$ \>  Global kinetic-energy components (KS, KZ and KE in Figures) \\
$\overline{k_{cS}}$, $\overline{k_{cZ}}$ \>  Complementary kinetic-energy components. \\
$K$  \>  Global kinetic energy \\ 
$L$  \>   Mixing length for the vertical dissipation scheme (Part~II) \\
$p$, $p_r$, $p_{00}$ \> Local pressure, reference and standard values of  $p$  \\
$p_t$, $p_b$ \> Pressure at top and bottom of atmosphere \\
$\dot{q}$ \>  Diabatic heating \\
$R$ \> Gas constant \\
${\cal R}$  \>   Earth radius \\
$s$, $s_r$, $s_{00}$  \> Local specific entropy, reference and standard values of $s$ \\
$S$  \>  Global entropy \\ 
$T$, $T_r$, $T_{00}$ \> Local temperature, reference and standard values of $T$ \\
$T_1$, $T_2$ \>  Two reference temperatures used in section 6 (Part~I) \\
``$T_0 \Sigma$''  \>   Global static entropic energy \\
TPE \> Global total potential energies (Lorenz, 1955) \\
${\bf U}_h = (u,v) $ \> Horizontal wind speed and its components \\
${\bf U}_g = (u_g,v_g) $ \> Geostrophic  horizontal wind and its components \\
${\bf U}_a= (u_a,v_a) $ \> Ageostrophic horizontal wind and its components \\
$V_a$, $V_b$ \> Geostrophic and ageostrophic wind in complex notation (Part~II) \\
$X$, $X_1$, $X_2$ \>  Dummy arguments of $\cal{F}$ and $\cal{G}$  exergy (quadratic) functions \\
$z$, $z_b$  \> Height above surface and bottom of atmosphere  (Part~II) \\
($a$, $b$, $c$)  \> Coefficients used in section 5  (Part~I) \\
$\ell$  \> Dummy value used in section 5  (Part~I) \\
${\alpha}$, ${\alpha}_r$ \> Inverse of density, reference value of ${\alpha}$ \\
$\eta$  \>  A dummy variable \\
${\varepsilon}_1$, ${\varepsilon}_2$  \> Two error functions \\
$\phi = g \: z$ \>  Local specific potential energy ($g$ is the acceleration due to gravity) \\
$\lambda$ \>  Longitude \\
$\varphi$ \>  Latitude \\
$\Lambda$ \>  A scale height (Part II) \\
$\overline{\sigma}\,(p)$ \> Average static stability on a pressure level (Lorenz, 1955) \\
$\omega = d/dt(p)$ \> Vertical velocity in pressure coordinates \\
$\Omega$  \>   Angular velocity of the Earth \\
$\kappa=R/c_p$ \>  A non-dimensional number \\
$\theta$ \>  A general surface angle (Part II) \\
${\bf \nabla}_{\!p}(\phi)$ \> Pressure force \\
$[{\bf \nabla}_{\!p}(\phi)]_x$, $[{\bf \nabla}_{\!p}(\phi)]_y$  \> Horizontal components of pressure force \\
$d/dt(\ldots)$ \> ${\partial}/{\partial t}(\ldots) + 
                             {{\bf U}_h} \: . \: {\bf \nabla}_{\!p}(\ldots) +
                             \omega \: {\partial}/{\partial p}(\ldots)$ :  the material derivative \\
${\cal V}_T$  \>   A non-dimensional number \\
$\psi$, $\Psi$  \>  Two dummy variables \\
\end{tabbing}

\vspace*{2mm}
The notation used in this paper is adapted from 
Reiter (1969). Let us consider any pressure level within 
a limited-area domain limited by
south and north latitudes ${\varphi}_s $
and ${\varphi}_n $ and by east and west 
longitudes ${\lambda}_e $ and ${\lambda}_w$.
Horizontal averaging operators ${\psi}^{\varphi}(\lambda, p)$ 
and ${\psi}^{\lambda}(\varphi, p)$ will be indicated by superscripts, 
they are defined for any local scalar ${\psi}(\lambda, \varphi, p)$ by
\vspace{-0.15cm}
\begin{eqnarray}
\hspace{-0.0cm}
    {\psi}^{\varphi}  \:& =& \:
     \frac{1}{\sin({\varphi}_n) - \sin({\varphi}_s)} \;
     \int_{{\varphi}_s }^{{\varphi}_n } \! {\psi} \: \cos(\varphi) \;d\varphi \: ,
     \nonumber \\
    {\psi}^{\lambda}  \: &=& \:
     \frac{1}{{\lambda}_e - {\lambda}_w} \;
     \int_{{\lambda}_w }^{{\lambda}_e } \! {\psi} \; d\lambda
\: . \nonumber
\end{eqnarray}

The average and departure terms are defined by
\vspace{-0.15cm}
\begin{eqnarray}
\hspace{-0.0cm}
     \overline{{\psi}}      \hspace{-0.2cm} & = & \hspace{-0.2cm}  
     ({{\psi}}^{\varphi}){}^{\lambda} \: =   \: 
     ({{\psi}}^{\lambda}){}^{\varphi} \: =   \:
        {{\psi}}^{\lambda}{}^{\varphi}
\: ,  \label{AppAdep1}  \\
          {\psi}_{\lambda}  \hspace{-0.2cm} & = & \hspace{-0.2cm} 
          {\psi} - {{\psi}}^{\lambda}
\; \;   ;   \; \;
          {\psi}^{\lambda}_{\varphi}     \:  =  \:
          {\psi}^{\lambda} - {{\psi}}^{\lambda}{}^{\varphi} \:  =  \:
          {\psi}^{\lambda} - \overline{{\psi}}
 \: ,  \label{AppAdep2}  \\
          {\psi}^{\:\prime}  \hspace{-0.2cm} & = & \hspace{-0.2cm} 
          {\psi} - \overline{{\psi}}     \:  =  \:
          {\psi}^{\lambda}_{\varphi} + {\psi}_{\lambda}
\: ,  \label{AppAdep3}
\end{eqnarray}
where departure terms are indicated by subscripts.

The global value ${\Psi}$ is defined from any local value 
${\psi}(t, \lambda, \varphi, p)$ by
\vspace{-0.15cm}
\begin{eqnarray}
\hspace{-0.0cm}
    {\Psi}  \: = \int_{ p_t }^{ p_b } 
                 \overline{{\psi}} \; \;  \frac{dp}{g}
\: , \nonumber
\end{eqnarray}
where $p_t$ and $p_b$ are the pressure at the bottom
and top of the atmosphere. 
The horizontal and time derivatives at constant pressure and
the vertical derivative are:
\vspace{-0.15cm}
\begin{eqnarray} 
{\partial}_x   \hspace{-0.2cm} & = & \hspace{-0.2cm} 
         \frac{\partial}{\partial x}\:(\ldots) \: = \:
          \frac{1}{ {\cal R} \cos(\varphi) }
       \: {\left[ \frac{\partial}{\partial \lambda}\: (\ldots) \right]}_{(t, {\varphi}, p)}
       \: ,  \nonumber  \\
{\partial}_y   \hspace{-0.2cm} & = & \hspace{-0.2cm}  
         \frac{\partial}{\partial y} \: (\ldots) \: = \:
         \frac{1}{\cal R} \: 
       \: {\left[ \frac{\partial}{\partial \varphi}\: (\ldots) \right]}_{(t, {\lambda}, p)}
       \: ,  \nonumber  \\
{\partial}_t   \hspace{-0.2cm} & = & \hspace{-0.2cm}  
         \frac{\partial}{\partial t} \: (\ldots) \: = \:
       \: {\left[ \frac{\partial}{\partial t}\: (\ldots) \right]}_{( {\lambda}, {\varphi}, p)}
       \: ,  \nonumber \\
{\partial}_p   \hspace{-0.2cm} & = & \hspace{-0.2cm}  
         \frac{\partial}{\partial p} \: (\ldots) \: = \:
      - \: \frac{R \: T }{g \: p } 
       \: {\left[ \frac{\partial}{\partial z}\: (\ldots) \right]}_{(t, {\lambda}, {\varphi})}
       \: .  \nonumber
\end{eqnarray}

\vspace{6mm}
\noindent
{\Large\bf Appendix B.    The local available enthalpy cycle.}
             \label{appendix_B}
\renewcommand{\theequation}{B.\arabic{equation}}
  \renewcommand{\thefigure}{B.\arabic{figure}}
   \renewcommand{\thetable}{B.\arabic{table}}
      \setcounter{equation}{0}
        \setcounter{figure}{0}
         \setcounter{table}{0}
\vspace*{1mm}
\hrule

\vspace*{2mm}
The first stages of the computations leading to the available 
enthalpy cycle (\ref{eq:cycle2}) will be described, though only
for the component $a_Z$. Similar methods can be applied
to the five other components. The first step is to compute the
derivation at constant pressure with respect to any $\eta$ variable 
($\eta=t$, $\lambda$, $\varphi$), or with respect to pressure
if $\eta = p$. The result is
\vspace{-0.15cm}
\begin{eqnarray}
    \overline{ {\partial}_{\eta} ( a_Z )}
    \: & = & \:
       \overline{ 
       {\partial}_{\eta}
             \left\{ \:
             c_p \: T_r \: {\cal F} \!
             \left(
             \frac{T^{\lambda}_{\varphi}}{\overline{T}} 
             \right)
             \: \right\}
       }
    \nonumber \\
    \: & = & \:
       c_p \: \left( \frac{T_r}{\overline{T}} \right) \: 
       \overline{ 
          \left\{ \:
             \: \left( \frac{\overline{T}}{T^{\lambda}} \right)  
             \: {T^{\lambda}_{\varphi}} \;
            {\partial}_{\eta} (T^{\lambda}) - {\partial}_{\eta} (\overline{T})
          \: \right\}
       }
       \: . \label{AppB:compdtaz1}
\end{eqnarray}
The time derivative is transformed using the commutating 
properties between ${\partial}_{t}(...)$ and 
${(...)}^{\lambda}$, to give
\vspace{-0.15cm}
\begin{eqnarray}
    \overline{ {\partial}_{t} ( a_Z )}
    \hspace{-0.2cm} & = & \hspace{-0.2cm}
       c_p \: \left( \frac{T_r}{\overline{T}} \right) \:
       \overline{ 
          \left\{ \:
             \:  \left( \frac{\overline{T}}{T^{\lambda}} \right)  \: 
                 {T^{\lambda}_{\varphi}} \;
             {({\partial}_{t} T)}^{\lambda} - \overline{({\partial}_{t} T)}
          \: \right\}
       }
       \: . \label{AppB:compdtaz2}
\end{eqnarray}
The boundary terms are obtained from (\ref{AppB:compdtaz1})
for $a_Z$ and with the  equivalent equation for $a_{cZ}$ given by 
(\ref{defacZ}). The results can be rearranged into
\vspace{-0.15cm}
\begin{eqnarray}
\hspace{-0.8cm}
    \overline{ B ( a_Z )}
    \hspace{-0.2cm} & = & \hspace{-0.2cm}
       c_p \: \left( \frac{T_r}{\overline{T}} \right) \:
       \overline{ 
          \left\{ \:
             \:  \left( \frac{\overline{T}}{T^{\lambda}} \right)  \:
                 {T^{\lambda}_{\varphi}} \;
             { B({T}^{\lambda}) } - { B(\overline{T}) }
          \: \right\}
       }
      \; , \label{AppB:compdtaz3} 
  \\
\hspace{-0.8cm}
    \overline{ B ( a_{cZ} )}
    \hspace{-0.2cm} & = & \hspace{-0.2cm}
       c_p \: \left( \frac{T_r}{\overline{T}} \right) \:
       \overline{ 
          \left[
             \left( \frac{T^{\lambda}_{\varphi}}{T^{\lambda}} \right)  \: { B({T}) }
          -
             \left( \frac{T_{\lambda}}{\overline{T}} \right)  \: { B(\overline{T}) }
          +
             \left\{ \frac{T \: {\overline{T}}}{ {( T^{\lambda})}^2 } -1 \right\}  \: B(T^{\lambda})
          \right]
       }
        \; . \label{AppB:compdtaz4}
\end{eqnarray}

Equation (\ref{defevolT}) in then used to express 
$\overline{ {\partial}_t ( T) } $ in 
(\ref{AppB:compdtaz2}) and,
after long and exact manipulations, the quantity 
$\overline{ {\partial}_t ( a_Z ) } +
 \overline{  B( a_Z + a_{cZ} ) } $,
which is equal to 
(\ref{AppB:compdtaz2})$+$
(\ref{AppB:compdtaz3})$+$
(\ref{AppB:compdtaz4}), is found to be
equal to the sum 
$\: + \:\overline{c_{AS}} - \overline{c_Z}
      - \overline{c_A}    + \overline{g_Z}$,
as indicated in (\ref{eq:cycle2}),
without approximations or cancelled terms.

\vspace{6mm}
\noindent
{\Large\bf Appendix C.   Approximation formulas for $g_Z$ and $g_E$.}
             \label{appendix_C}
\renewcommand{\theequation}{C.\arabic{equation}}
  \renewcommand{\thefigure}{C.\arabic{figure}}
   \renewcommand{\thetable}{C.\arabic{table}}
      \setcounter{equation}{0}
        \setcounter{figure}{0}
         \setcounter{table}{0}
\vspace{1mm}
\hrule

\vspace*{2mm}
Starting from Eqs.~(\ref{eq:ets94}) and (\ref{eq:ets95}),
the diabatic heating $(\dot{q})$ is separated
differently for $\overline{g_Z}$ and $\overline{g_E}$.
It is found, with the use of
${[ T^{\lambda}_{\varphi} \: {(\dot{q})}_{\lambda} ]}^{\lambda}=0$
for $\overline{g_Z}$, that
\vspace*{-0.15cm}
\begin{eqnarray}
\frac{\overline{g_Z}}{T_r} \: & = & \:
\overline{ 
  \left[ \: 
   \left( \frac{T^{\lambda}_{\varphi}}{\overline{T}} \right)
     \; \left( 
        \frac{ \: \overline{(\dot{q})} 
             + {(\dot{q})}^{\lambda}_{\varphi} 
             + {(\dot{q})}_{\lambda}
        }{{T}^{\lambda}}
    \: \right)
  \right]
} \; ,
\nonumber \\
  \: &=& \:
\overline{\left(
        \frac{T^{\lambda}_{\varphi}}{T^{\lambda}} 
               \right)}
\; \;
    \frac{ \overline{(\dot{q})} }{\overline{T} }
\: + \:
\overline{
 \left[ \; 
   \: \frac{T^{\lambda}_{\varphi}}{{T}^{\lambda}} \;
   \; \frac{  {(\dot{q})}^{\lambda}_{\varphi} }{\overline{T}}
\; \right]
}
\: , \nonumber 
\end{eqnarray}
and
\vspace*{-0.15cm}
\begin{eqnarray}
\frac{\overline{g_E}}{T_r} \: & = & \:
\overline{ 
  \left[ \:
    \left( \frac{T_{\lambda} }{ T} \right)
     \; \left( \: 
      \frac{
                {(\dot{q})}^{\lambda}
             + {(\dot{q})}_{\lambda}
               }{{T}^{\lambda}}
     \right)
  \right]
} \: ,
\nonumber \\
 \: &=& \:
\overline{ 
  \left[ \: {\left( \frac{T_{\lambda}}{T} \right)}^{\lambda}
     \; \frac{{(\dot{q})}^{\lambda} }{ {T}^{\lambda} }
  \right]
}
\: + \:
\overline{ 
   \left[ \:
     \left(
     \frac{T_{\lambda}}{{T}^{\lambda}}
     \right)
     { 
        \left(
      \frac{{(\dot{q})}_{\lambda}}{T} 
        \right)
     }^{\lambda}
 \right]
}
\: . \nonumber
\end{eqnarray}

The last terms of these equations, say 
$\overline{ \: [ T^{\lambda}_{\varphi} \; 
 {(\dot{q})}^{\lambda}_{\varphi} / {T}^{\lambda} ]}$
and
$\overline{ \{ \: {[ T_{\lambda} \: 
 {(\dot{q})}_{\lambda} / T ]}^{\lambda}
 / {T}^{\lambda}\} }$,
are close to the definition given in P78
and it can be demonstrated that the first
terms
$\overline{( T^{\lambda}_{\varphi} / T^{\lambda} )}
 \; [ \overline{(\dot{q})} / \overline{T} ]
$
and
$\overline{ \{ \: {( T_{\lambda} / T )}^{\lambda}
 \; [ {(\dot{q})}^{\lambda} / {T}^{\lambda} ] \} }$
are one order of magnitude smaller.
Indeed, it appears that absolute
values of the first non dimensional terms
$A = \overline{( T^{\lambda}_{\varphi}/T^{\lambda} )}$
and $B = {( T_{\lambda} / T )}{}^{\lambda}$ are small 
when compared to unity. The demonstration starts with
$A = 1-\overline{(\overline{T}/T^{\lambda})}$ and
$B = 1-{( T^{\lambda} / T )}{}^{\lambda}$.
From Eq.~(\ref{AppAdep2}), the use of 
$T^{\lambda} = T^{\lambda}_{\varphi} + \overline{T}$
in $A$ and of $T = T^{\lambda} + T_{\lambda}$ in $B$
lead to 
\begin{equation}
A = 1 - \overline{ 
      \left( 
           \frac{1 }{ 1 + T^{\lambda}_{\varphi} / \overline{T} }
      \right) 
      }
\; \; \; \; \; \mbox{and} \; \; \; \;
B = 1 -
          \left( 
               \frac{ 1 }{ 1 +  T_{\lambda} / T^{\lambda} } 
          \right)^{\lambda} \: .
\nonumber
\end{equation}

For small $|x|$, $1/(1+x) \approx
1 - x + x^2$ and the limits for small
$|T^{\lambda}_{\varphi} / \overline{T}|$ and
$|T_{\lambda} / T^{\lambda}|$ are 
$A \approx  
-\overline{\{ {(T^{\lambda}_{\varphi} / \overline{T})}{}^2 \}}$ and
$B \approx  -{ \{ {(T_{\lambda} / T^{\lambda})}{}^2 \}}{}^{\lambda}$.
The first order term ``$-x$'' cancels out in both cases
because $\overline{(T^{\lambda}_{\varphi})} 
= {(T_{\lambda})}^{\lambda} = 0$.
As a consequence, the terms
$\overline{ \{ \: {( T_{\lambda} / T )}^{\lambda}
 \; [ {(\dot{q})}^{\lambda} / {T}^{\lambda} ] \} }$
and
$\overline{( T^{\lambda}_{\varphi} / T^{\lambda} )}
 \; [ \overline{(\dot{q})} / \overline{T} ]$
in the expressions above are small and the following 
approximations 
(\ref{AppC:etsGZapprox}) and (\ref{AppC:etsGEapprox}) are
close to the results obtained in P78, namely
$\overline{ \{
{T}^{\lambda}_{\varphi} \: {(\dot{q})}^{\lambda}_{\varphi} 
\} } / T_r$   and 
$\overline{ \{ {T}_{\lambda} \: {(\dot{q})}_{\lambda}
\} } / T_r$, 
respectively. 
The local available-enthalpy versions of generations terms thus become
\vspace*{-0.15cm}
\begin{eqnarray}
\overline{g_Z}  \! \! & \approx & \! \!
 \overline{
   \: \left\{ 
     \frac{
      T^{\lambda}_{\varphi}
      \; {(\dot{q})}^{\lambda}_{\varphi}
      }
     {{T}^{\lambda}} 
     \right\}
 } 
 \: \left( \frac{T_r}{\overline{T}} \right)
\: , \label{AppC:etsGZapprox} 
\\
\overline{g_E} \! \! & \approx & \! \!
\overline{ 
   \left\{ \: 
     { \left\{ 
     \frac{T_{\lambda} \: {(\dot{q})}_{\lambda}}
      {T}
      \right\} 
      }^{\lambda}
     \: \left( \frac{T_r}{{T}^{\lambda}} \right) 
  \right\}
 }
\: . \label{AppC:etsGEapprox}
\end{eqnarray}
They correspond to equations mentioned at the end of section 4.


\vspace{5mm}
\noindent{\Large\bf \underline{References}.}
\vspace{2mm}

\noindent{$\bullet$ Andrews,~D.} {1981}.
{A note on potential energy in a stratified compressible fluid.
{\it J. Fluid Mech.\/},
{\bf 107,} 
p.227--236.}

\noindent{$\bullet$ Bernard-Bouissi\`eres,~I., 
Cadiou,~M., Muzellec,~A., Vincent,~Ch.} {1991}.
Cycles \'energ\'etiques.
{\it Internal report of the French School of Meteorology.\/}

\noindent{$\bullet$ Brennan,~F.~E. and Vincent,~D.~G.} {1980}.
{Zonal and eddy components of the synoptic-scale energy
budget during intensification of hurricane Carmen (1974).}
{\it Mon. Weather Rev.\/}
{\bf 108,}
p.954--965.

\noindent{$\bullet$ Carnot,~N,~L,~S.} {1824}.
{R\'eflexions sur la puissance motrice du feu, et sur les machines propres \`a d\'evelopper cette puissance.
See the account of Carnot's theory written by W. Thomson (1849) in the
{\it Trans. Roy. Soc. Edinb.\/}
{\bf 16},
p.541--574.
An English translation by R. H. Thurston of the version published in the ``Anales scientifique de l'\'Ecole Normale Sup\'erieure'' (ii. series, t.1, 1872) is available in the url: \url{http://www3.nd.edu/~powers/ame.20231/carnot1897.pdf} (Wiley \& Sons, 1897, digitized by Google)} 

\noindent{$\bullet$ Clausius,~R.} {1865}.
{\"Uber verschiedene f\"ur die Anwendung bequeme Formen der
Hauptgleichungen der mechanischen W\"armetheorie.
(On Different Forms of the Fundamental Equations of the Mechanical Theory of Heat).
{\it Ann. der Phys. und Chem.\/}
{\bf 125},
p.353-400.}

\noindent{$\bullet$ Courtier,~J.~A., 
Freydier,~C., Geleyn,~J.F., 
Rabier,~F. and Rochas,~M.} {1991}.
{The Arp\`ege project at M\'et\'eo-France.
{\it ECMWF Seminar Proceedings.\/},
Reading, 9-13 Sept. 1991, Volume II,
p.193--231.}

\noindent{$\bullet$ Dutton,~J.~A.} {1973}.
{The global thermodynamics of atmospheric motion.
{\it Tellus.\/}
{\bf 25,} (2),
p.89--110.}

\noindent{$\bullet$ Engel,~F., Petit,~B., Pontaud,~M.} {1992}.
Un cycles \'energ\'etiques local associ\'e au mod\`ele 
non-hydrostatique de COME : applications \`a une 
onde d'est.
{\it Internal report of the French School of Meteorology.\/}

\noindent{$\bullet$ Gibbs,~J.~W.} {1873}.
{A method of geometrical representation
of the thermodynamic properties of substance
by means of surfaces.
{\it Trans. Connecticut Acad.\/}
{\bf II}: p.382--404.
(Pp 33--54 in Vol. 1 of
{\it The collected works of J. W. Gibbs,\/}
1928.
Longmans Green and Co.)} 

\noindent{$\bullet$ Karlsson,~S.} {1990}.
{{\it Energy, Entropy and Exergy in the atmosphere.\/}
Thesis of the Institute of Physical Resource Theory.
Chalmers University of Technology.
 G\"oteborg, Sweden.}

\noindent{$\bullet$ Kasahara,~A.} {1974}.
{Various vertical coordinate systems used for numerical 
weather prediction.
{\it Mon. Weather Rev.\/}
{\bf 102},
p.509--522.}

\noindent{$\bullet$ Kucharski,~F.} {1997}.
{On the concept of exergy and available potential energy.
{\it Q. J. R. Meteorol. Soc.\/}
{\bf 123},
p.2141--2156.} 

\noindent{$\bullet$ Livezey,~R.~E. and Dutton,~J.~A.} {1976}.
{The entropic energy of geophysical fluid systems.
{\it Tellus.\/}
{\bf 28,} (2),
p.138--157.} 

\noindent{$\bullet$ Lorenz,~E.~N.} {1955}.
{Available potential energy and the 
 maintenance of the general circulation.
{\it Tellus.\/}
{\bf 7,} (2),
p.157--167.}

\noindent{$\bullet$ McHall, ~Y.~L.}, {1990}.
{Available potential energy in the atmospheres.
{\it Meteorol. Atmos. Phys.\/}, 
{\bf 42}, 
p.39--55.}

\noindent{$\bullet$ Margules,~M.} {1903-05}.
{On the energy of storms. 
{\it Smithsonian Miscellaneous collections,} 
51, 4, 533--595, 1910. 
{\it (Translation by C. Abbe from the appendix to the annual 
volume for 1903 of the Imperial Central Institute for Meteorology, 
Vienna, 1905.
`\"Uber die energie der st\"urme'. 
{\it Jahrb. Zentralantst. Meteorol.},
{\bf 40,} p.1--26, 1903)}.}

\noindent{$\bullet$ Marquet~P.} {1991}.
{On the concept of exergy and available
enthalpy: application to atmospheric energetics.
{\it Q. J. R. Meteorol. Soc.}
{\bf 117}:
p.449--475.
\url{http://arxiv.org/abs/1402.4610}.
{\tt arXiv:1402.4610 [ao-ph]}}

\noindent{$\bullet$ Marquet,~P.} {1994}.
{\it Applications du concept d'exergie \`a
l'\'energ\'etique de l'at\-mosph\`ere. Les
notions d'enthalpie utilisables s\`eche
et humide\/}.
PhD-thesis of the Paul Sabatier University.
Toulouse, France.

\noindent{$\bullet$ Marquet,~P.} {1995}.
{On the concept of pseudo-energy of T. G. Shepherd.
{\it Q. J. R. Meteorol. Soc.\/}
{\bf 121}:
p.455--459.
\url{http://arxiv.org/abs/1402.5637}.
{\tt arXiv:1402.5637 [ao-ph]}}

\noindent{$\bullet$ Marquet,~P.} {2001}.
{\it The available enthalpy cycle. 
 Applications to idealized baroclinic 
 waves.\/}.
Note de centre du CNRM. Number 76.
Toulouse, France.

\noindent{$\bullet$ Marquet~P.} {2003a}.
{The available enthalpy cycle. Part~I : 
 Introduction and basic equations.
{\it Q. J. R. Meteorol. Soc.\/},
{\bf 129}, (593),
p.2445--2466.}

\noindent{$\bullet$ Marquet~P.} {2003b}.
{The available enthalpy cycle. Part~II : 
 Applications to idealized baroclinic waves.
{\it Q. J. R. Meteorol. Soc.\/},
{\bf 129}, (593),
p.2467--2494.}

\noindent{$\bullet$ Maxwell,~J.~C.} {1871}.
{Theory of Heat. References are made in the text to next editions of this book. 
{\it Longmans, Green and Co.}
London.}

\noindent{$\bullet$ Michaelides,~S.~C.} {1987}.
{Limited area energetics of Genoa cyclogenesis.
{\it Mon. Weather Rev.\/}
{\bf 115},
p.13--26.}

\noindent{$\bullet$ Muench,~H.~S.} {1965}.
{On the dynamics of the wintertime stratosphere 
circulation.
{\it J. Atmos. Sci.\/}
{\bf 22},
p.349-360.}

\noindent{$\bullet$ Normand,~Sir~C.} {1946}.
{Energy in the atmosphere.
{\it Q. J. R. Meteorol. Soc.\/},
{\bf 72,}
p.145--167.}

\noindent{$\bullet$ Pearce,~R.~P.} {1978}.
{On the concept of available potential energy.
{\it Q. J. R. Meteorol. Soc.\/}
{\bf 104},
p.737--755.}

\noindent{$\bullet$ Pichler,~H.} {1977}.
{Die bilanzgleichung f\"ur die statischer 
entropische Energie der Atmos\-ph\"are.
{\it Arch. Met. Geoph. Biokl.\/}, 
Ser.A, 
{\bf 26},
p.341--347.}

\noindent{$\bullet$ Reiter,~E.~R.} {1969}.
{Mean and eddy motions in the atmosphere.
{\it Mon. Wea\-ther Rev.\/},
{\bf 97,}
p.200--204.\/}

\noindent{$\bullet$ Saltzman,~B. and Fleischer,~A.} {1960}.
{The modes of release of available potential energy 
  in the atmosphere.
{\it J. Geophys. Res.\/}
{\bf 65}, (4),
p.1215--1222.}

\noindent{$\bullet$ Shepherd,~T.~G.} {1993}
{A unified theory of available potential energy.
{\it Atmosphere-Ocean.\/},
{\bf 31,}
p.1--26.}

\noindent{$\bullet$ Thomson,~W.} {1849}.
{An account of Carnot's theory of the ``Motive Power of Heat'',
 with numerical results deduced from Regnault's experiments on 
steam. 
{\it Trans. Roy. Soc. Edinb.\/}
{\bf 16,} Part~5,
p.541--574.}

\noindent{$\bullet$ Thomson,~W.} {1853}.
{On the restoration of mechanical energy 
from an unequally heated space.
{\it Phil. Mag.\/}
{\bf 5,} 30, 4e series,
p.102--105.}

\noindent{$\bullet$ Thomson,~W.} {1879}.
{On thermodynamic motivity. 
{\it Phil. Mag.\/}
{\bf 7,} 44, 5e series,
p.346--352.} 

 \end{document}